\documentclass[fleqn,10pt]{wlscirep}
\usepackage[utf8]{inputenc}
\usepackage[T1]{fontenc}
\usepackage{subcaption}

\usepackage{rotating}

\title{First-mover advantage explains gender disparities in physics citations}

\author[1]{Hyunsik Kong}
\author[1]{Samuel Martin-Gutierrez}
\author[1,*]{Fariba Karimi}
\affil[1]{Networked Inequality group, Complexity Science Hub, Josefstaedter Strasse 39, Vienna, 1080, Austria}

\affil[*]{corresponding author: karimi@csh.ac.at}

\begin{abstract}
Mounting evidence suggests that publications and citations of scholars in the STEM fields (Science, Technology, Engineering and Mathematics) suffer from gender biases. In this paper, we study the physics community, a core STEM field in which women are still largely underrepresented and where these gender disparities persist. To reveal such inequalities, we compare the citations received by papers led by men and women that cover the same topics in a comparable way. To do that, we devise a robust statistical measure of similarity between publications that enables us to detect pairs of similar papers. Our findings indicate that although papers written by women tend to have lower visibility in the citation network, pairs of similar papers written by men and women receive comparable attention when corrected for the time of publication. These analyses suggest that gender disparity is closely related to the first-mover and cumulative advantage that men have in physics, and is not an intentional act of discrimination towards women.
\end{abstract}
\begin{document}

\flushbottom
\maketitle
%
%
\thispagestyle{empty}


\section*{Introduction}

Mounting evidence suggests gender bias in publications and citations of scholars in STEM \cite{caplar2017quantitative,Dworkin2020}. Such biases can result in situations where women (or other under-represented minorities) may feel invisible and ignored in male-dominated environments. The feeling of not being part of the community can result in a higher dropout rate among women, a phenomenon known as leaky pipeline \cite{alper1993pipeline}. Leakage in the academic pipeline consequently affects the academic community for generations to come due to a lack of diversity, inclusion, innovation and role models. Thus, it is of utmost societal importance to accurately identify those biases and devise bottom-up approaches to tackle them.

Gender inequality in academia manifests itself in the production of science and performance outcomes. Some inequalities are inevitable: parenthood, career breaks, limited access to role models and resources can create situations in which women and other minorities show less productivity and performance compared with their white male peers. Frequently, these unavoidable inequalities are exacerbated through formal and informal social relationships, which in turn affect the citation network structure and reinforce existing inequalities.



Academic productivity is often associated with number of publications 
throughout a researcher's career. Previous studies have found that women publish fewer peer-reviewed articles than men \cite{kaufman2011gender,reed2011gender}, while a more recent study found that the disparity in the productivity of men and women disappears if we compare the productivity with regard to the scholar's career length \cite{jadidi2018gender,huang2020historical}. Women display higher publication rates later in their academic careers, but take up fewer leadership roles \cite{maske2003determinants, reed2011gender}. Mueller et al. suggest that publication productivity may be a factor that hinders women from advancing within surgery \cite{mueller2017publication}, while Reed and colleagues point out that mid-career assessment of productivity may not be an appropriate measure of leadership skills \cite{reed2011gender}. 




Beyond disparities in publication and productivity, analysing citation patterns can help to identify whether gender differences exist in the way scholars award and recognize each others' works. In other words, while productivity is associated with individual or collaborative efforts, citation is an indication of how these efforts are perceived by the community \cite{barabasi2018formula}. In this sense, one can argue that while the former operates among a small number of collaborators, the latter is related to the social processes that govern the community of scholars at large.


Previous studies have shown that patterns of citation can be different for men and women. This could be explained by intentional decision, quality difference, or paradigmatic research topics \cite{aksnes2011female, lindsey1989using, davenport1995cites}. It has been argued that in the most productive countries, all articles with women in key author positions receive fewer citations than those with men in the same positions \cite{https://doi.org/10.1002/asi.22784, Lariviere2013}. 
Moreover, some research concluded that the differences in citation rates between men and women increase with the number of authors per article \cite{bendels2018gender}. This indicates that women are not only relatively less represented as high-impact key authors, but also that they attract significantly fewer citations for those key positions compared to men. One plausible assumption is that the lack of women in leadership positions causes this accentuated female under-representation (structural reasons) since the distribution of key authorships follows, by convention, a hierarchical order. In a recent paper, Dworkin and colleagues present a case study of citation patterns in top neuroscience journals, finding that papers for which first and last authors are men are over-represented in reference lists, and that the discrepancy is most prominent in the citation behaviours of men and is getting worse over time \cite{Dworkin2020}.

A major methodological obstacle is that simply comparing number of publications and citations of men and women is misleading. Men and women have different rates of entry in the scientific community for historical reasons, and when combined with other non-academic responsibilities, they may not show a similar behaviour at the aggregate level. Indeed, recent findings show that when differences in career length are controlled for, male and female scientists have similar rates of publication and citation on average \cite{huang2020historical}. However, beyond these insights on a population level, do men and women really receive different recognition for a similar work published around the same time? To truly examine the gender differences in citation, one should compare pairs of papers that cover the same topics in a comparable way. Relying on analysing only the average performance may hide variations that exist in data, and drive the community to inaccurate conclusions or inappropriate policies.


In this paper, we focus on analysing publication and citation patterns in the physics community as one of the core STEM areas where women are exceedingly under-represented, often facing belittling remarks and harassment \cite{barthelemy2016gender,aycock2019sexual}. More importantly, we examine gender differences not only at the population level, but also at the microscale by comparing pairs of statistically validated similar papers. 




\section*{Results}

We start by describing the dataset we have analysed and briefly explaining the methodology we have used to build the citation network and the pairs of similar papers. Then, we proceed to study gender disparities, first at the aggregate level and then by comparing pairs of similar papers.

\subsection*{Data Description}
\begin{figure}[ht]
     \centering
        \includegraphics[width=0.9\textwidth]{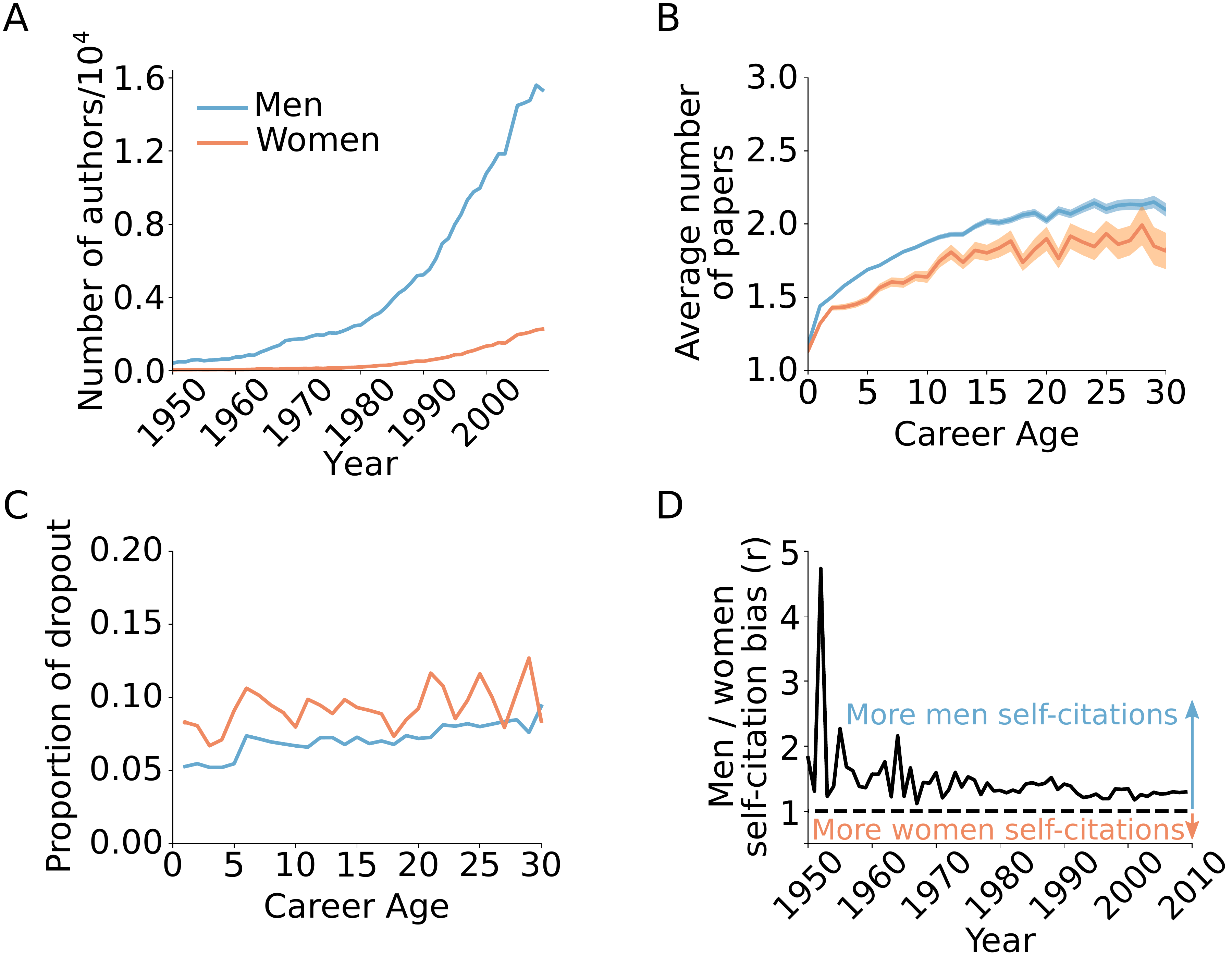}
        \caption{\textbf{Rate of growth of female participation, average publications by career age, dropout rate and annual ratio of male / female self-citations.} A: Number of male and female authors per year. B: Average number of publications by authors’ career age. C: Proportion of male and female authors who drop out compared to the remaining active authors per career age. D: Normalized ratio of male / female self-citations computed from (\ref{eq:r}) during the time period of interest. The horizontal dashed line is the line of equilibrium; data points above the equilibrium line indicate a higher ratio of male self-citation, and points below the line imply a higher ratio of female self-citation.}
        \label{fig:overview}
    \end{figure}

We study an American Physical Society (APS) dataset from 1893 to 2010 which contains articles' metadata, the authors' basic information, and the citations within the papers. The metadata consists of authors' full names and a unique digital object identifier (DOI) of the publication in a string format. 
For those names that are repeated in the dataset, we used name disambiguation methods proposed by Sinatra et al. to detect unique authors and correctly match authors to publications \cite{sinatra2016quantifying}. To infer gender from names, we implemented a gender-detection procedure that combines author names with an image-based gender inference technique applied to search results from Google Images \cite{karimi2016inferring}. This combined method results in high accuracy 
in the gender identification of scholars from different nationalities (see Supplementary Sections 1 and 2). 
The final dataset consists of 541,448 scholarly articles published over the course of 117 years. We have identified 70,833 gendered names, 9,947 women and 60,886 men. The evolution of the number of authors per year is shown in Figure \ref{fig:overview}A.

Here, the notion of "gender'' refers neither to the sex of the authors nor to the gender that the author self-identifies as. By the words "woman" or "female author", we mean an author whose name has a high probability of being assigned to female at birth or being identified as a woman due to facial characteristics. Given this limitation, we can safely argue that these methodologies are in accordance with social constructs and what people perceive as gender in society.





 


\subsection*{Constructing citation networks and assessing similar pairs}


We build the citation networks by considering each paper as a node and making a link from paper $i$ to paper $j$ if $i$ includes a citation to $j$. We measure the similarity between two papers using the \emph{bibliographic coupling strength} \cite{kessler1963bibliographic, egghe2002co}; that is, the number of publications that both papers cite. Two papers that cover similar topics in a comparable way are assumed to include a similar set of outgoing citations. However, within subfields there is usually a handful of classic publications that are cited in most works, so their inclusion in two different papers may not indicate actual similarity, but a \emph{citation convention}. To avoid such shortcomings of naive bibliographic coupling, and guarantee the significance of the overlapping set of citations, we apply a statistical test based on the hypergeometric distribution. This test controls for the incoming citations of the commonly cited papers and checks whether the size of the common set of citations is so large that it cannot be explained by randomness.

\begin{figure}[ht]
        \centering
        \includegraphics[width=0.9\textwidth]{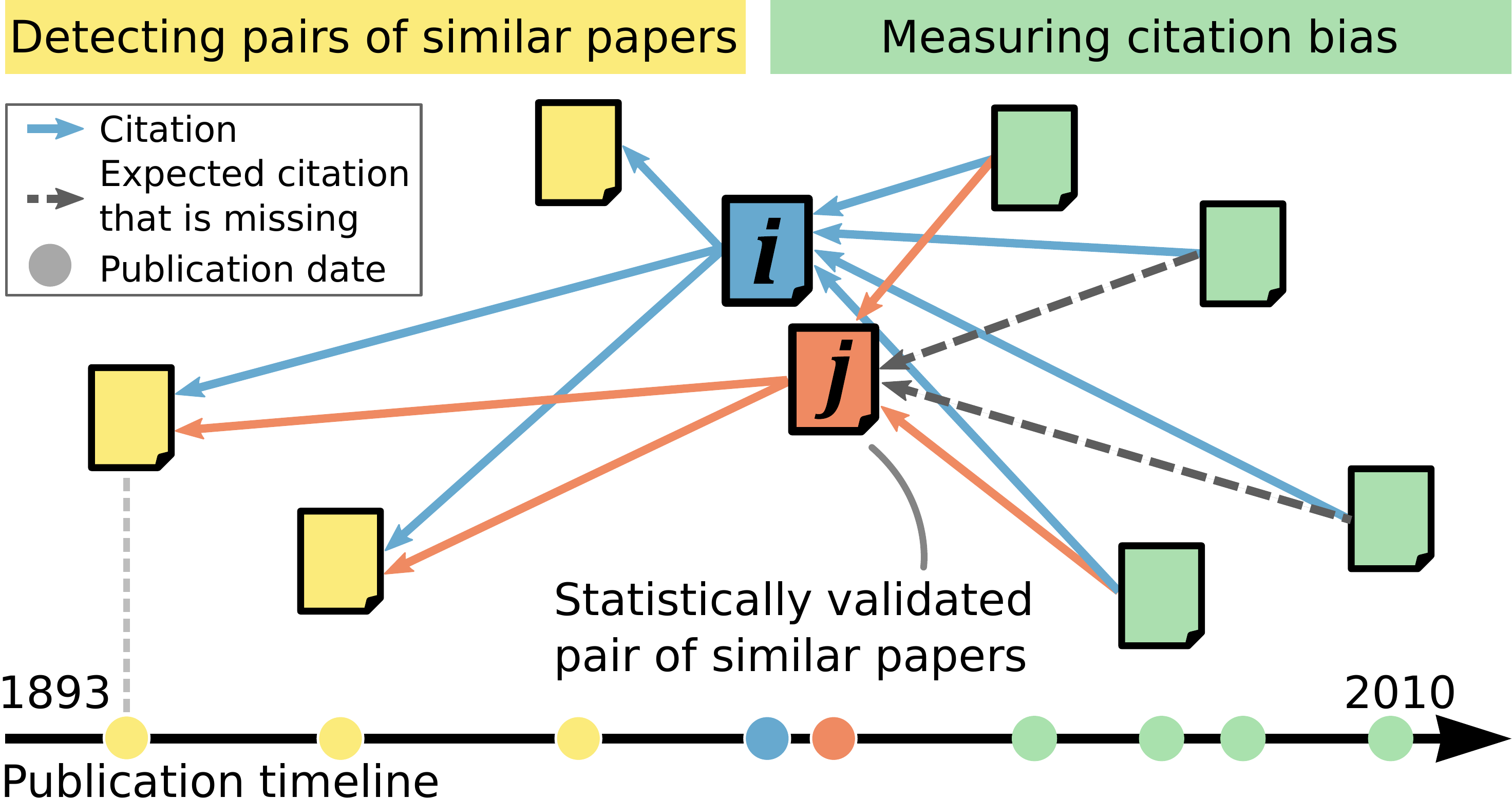}
        \caption{\textbf{Assessing similar pairs.} We use bibliographic coupling and hypergeometric statistical tests to select couples of similar papers based on their outgoing citation activities. Then we compare their respective popularity (incoming citations). Each node and each arrow represent a paper and a citation respectively, whereas each dashed arrow represents a potential citation that is missing.}
    \label{fig:fig1}
    \end{figure}

To explore gender disparities we select pairs of similar papers respectively written by male and female primary authors. Then, we compare the incoming citations to each element of the pair, such that, since the two publications are respectively led by a man and a woman, this comparison allows us to detect potential inequalities in the citation patterns. We have summarized this methodology in the diagram of Figure \ref{fig:fig1} and provided all the technical details in Methods.

\subsection*{Aggregate gender disparity trends}

To characterize the gender disparities at the aggregate level, we first analyse the aspects of scientific production that depend primarily on individual choices and ability: in particular, productivity, dropout rate, and self-citations. Then, we discuss authorship order, which depends on the internal organization of research groups. Finally, we study the behaviour of the scientific community as a whole by comparing the citations received by men and women.

\paragraph{Productivity.}
We define productivity as the number of publications that scholars produce during their career. In physics, we observe that women have a lower average number of publications compared to men across all their career ages (Figure \ref{fig:overview}B). While in the first two years of author's career the publication gap is closing, we observe a sudden increase in the gap from the second to the eleventh year. After this point, the publication gap starts decreasing again. These fluctuations in publication productivity can be associated with the disproportionate family responsibilities that women have to take on compared with men \cite{morgan2021unequal}. 

Although a researcher's productivity can be considered to be determined mainly by individual skills, the collaborative nature of scientific work makes it dependent on external factors such as other team members or departmental organization. Likewise, these factors, together with other aspects like social perception or family responsibilities, affect women's motivation to keep working in academia, potentially leading to the \emph{leaky pipeline} phenomenon. To quantify this phenomenon, in the next section we explore the difference in dropout rates between men and women. 


\paragraph{Dropout rate.}
We compute \textit{dropout} as a lack of publication activity for at least five years to distinguish the authors who are active in publishing from those who have dropped out. We investigate the ratio of dropout scholars at each career age compared to the number of active scholars by gender.
Figure \ref{fig:overview}C shows that female authors have a higher dropout ratio throughout their whole career. Most notably, the largest gaps appear in the early career years, with a 3.63\% difference between men and women in the fifth year. The dropout fluctuations after career age of 20 for women are caused by the low number of senior female scientists in the data (see Supplementary Figure \ref{fig:cacount}).  
The dropout rates of authors who leave academia after their first year (career age 0) are not shown in Figure \ref{fig:overview}C. This career age presents the highest dropout rates, with 28\% for male authors and 38\% for female authors.



\paragraph{Self-citation.}
Self-citation refers to cases where authors cite their own previous works. Self-citations increase the total citation count and the visibility of scholars \cite{king2017men, fowler2007does, maliniak2013gender}, potentially enhancing academic promotion and attention. We have measured the relative number of self-citations by all male and female authors with the following metric ($r$) to study the difference in self-citation ratios between the two genders over time \cite{king2017men}:


\begin{equation}
    r = \frac{\frac{\text{\% male self-citations}}{\text{\% male citations}}}{\frac{\text{\% female self-citations}}{\text{\% female citations}}}
    \label{eq:r}
\end{equation}

Figure \ref{fig:overview}D shows the temporal evolution of the ratio $r$. This result shows that women tend to cite themselves less than men and that this trend is consistent over the years (See Supplementary Table \ref{table:selfcit} for more details). 
Consequently, women's visibility in the citation network is partly penalized by the higher ratio of men citing their own previous works.


Another fundamental factor that affects an author's visibility is the position in which her name appears in the list of authors. This position depends on how the whole research group is organized and, crucially, in most cases it depends on the perceived level of contribution of each collaborator.


\paragraph{Authorship order analysis.}
In the majority of the scientific fields, including physics, the authorship order indicates relative contribution and seniority by putting emphasis on the first, the last, and the second positions \cite{baerlocher2007meaning, sauermann2017authorship}. In order to compare  the positions of authors, we first discard those papers for which authorship order is alphabetical. For this purpose, we perform a string comparison of the last names of the contributing authors and consider them to be in alphabetical order if the paper has at least four authors and all of them follow this order. Around 4\% of the papers can be considered as alphabetically ordered; in Supplementary Table \ref{table:ao} we detail their fraction by PACS subfield. After discarding those papers from the analysis, we study the authorship order in each publication and compare the proportion of female and male primary authors with what we would expect from the size of the population by conducting a two-proportion $z$-test (see Methods). 






The results show that male authors are listed in the first position of physics publications more frequently than expected
(See Supplementary Table \ref{table:aa}). 
We verified the robustness of this result by performing the same computation for last authors, obtaining analogous results. This is in line with previous findings that women feature only rarely as first or last authors in leading journals \cite{shen2018too}.

While the authorship order reflects how a researcher's coworkers perceive her contribution, the collective perception of the scientific community regarding the relevance of a paper is manifested in the citations of papers. In the following sections we will thoroughly compare the relative popularity of publications led by women and men.

\paragraph{Citation centrality analysis.}
The flow of citations determines the visibility and recognition of papers both locally and globally. To measure the local influence of papers we use the in-degree metric, and to measure the global influence, we use the PageRank centrality. Our aim is to verify if the visibility of papers written by women is proportionate to what we expect from their overall population size. To do that, we focus on the ranking of the nodes according to their respective centrality. 

Understanding ranking centrality is important for three reasons. First, the authors of papers in top ranks gain more visibility for themselves and those central papers influence future citation patterns \cite{bloch2017centrality, ding2009pagerank,karimi2018homophily}. Second, the visibility of papers in top ranks is being exacerbated by algorithmic tools such as Google Scholar. Third, since citation networks follow a heavy-tail distribution, those in top ranks stabilize their ranking position and give few opportunities for other papers to catch up \cite{ghoshal2011ranking}. Because of these network effects, it is important to study how minorities are represented in top ranks.

\begin{figure}[ht]
     \centering
    \includegraphics[width=1.0\textwidth]{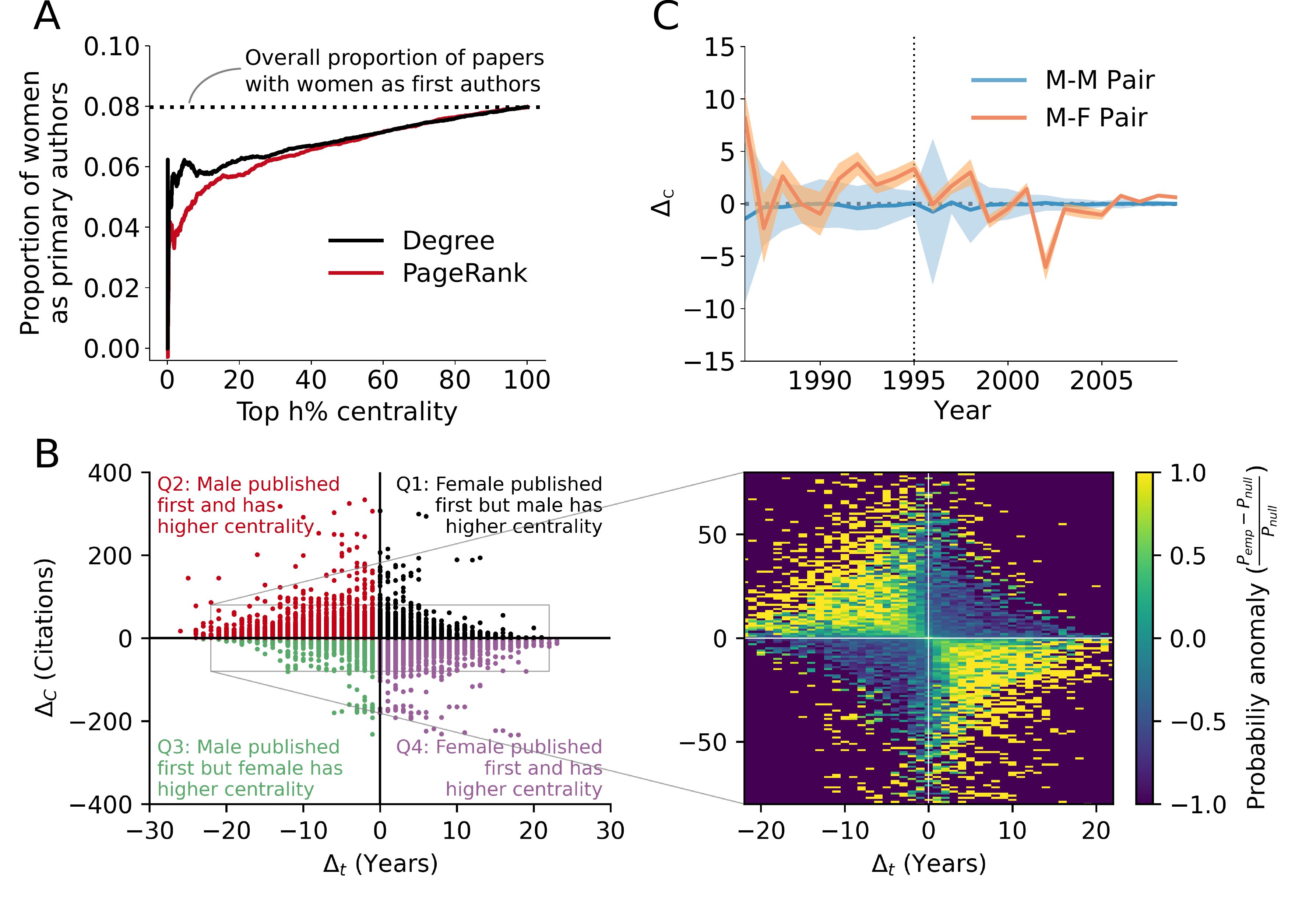}
    \caption{\textbf{Female author proportions in degree and PageRank centrality, evolution of centrality difference by year and relationship between time of publication and citation.} A: Proportion of publications with a female primary author per top $h$\% of degree (black) and PageRank centrality (red). The dotted horizontal line signifies the proportion of female primary authors in the observed samples. B Left: Citation and temporal differences between male-female pairs of papers with validated similarity. B Right: Heat map showing the probability anomaly of the joint probability distribution of citation and temporal differences computed with \eqref{eq:prob_anom}. C: Centrality differences of similar male-male pairs and similar male-female pairs over the years. The two papers within each pair are published no more than three years apart, and the publication year of the pair is defined by the year of the latter paper. The lines are the mean values and the shaded areas the standard errors.}
    \label{fig:centper}
    \end{figure}

We assigned each paper a gender by labeling it based on its first author. Then, we analysed the top $h$\% in-degree/PageRank centrality of the papers. Figure \ref{fig:centper}A suggests that papers written by women have notably lower in-degree and pagerank centrality than expected from their overall proportion. Female-led publications are substantially under-represented in the highest 
20th, 30th, and 40th percentages, and the deviation between the observed and the expected proportions likewise increases in the highest rank positions. While in-degree and PageRank follow a similar trend as expected, the proportion of females with high PageRank centrality is even lower when compared to the in-degree centrality. This suggests not only that papers written by women receive less attention but also that they are disadvantaged in terms of their position within the citation network.



So far, the global gender analysis points towards a notable disparity in productivity and citation of men and women. This could be partly due to historical reasons, to the cumulative advantage that early arrival confers to men, as well as to the high dropout rate of women \cite{huang2020historical}. The slower rate of arrival of women (see Figure \ref{fig:overview}A) may also play a relevant role. Together, these factors affect women's global visibility. The question that arises from these global results is, are scholars intentionally ignoring (and therefore, under-citing) research works led by women? To explore this possibility, in the following section we study pairs of papers written by men and women that are statistically validated twins, and measure the citations that each receives.






\subsection*{Pair-wise citation analysis}



We identified statistically validated male-female pairs of similar papers using the methodology described in Methods and summarized in Supplementary Figure \ref{fig:similarity_alg}.
Then, we computed the difference in the number of citations each member of the pair receives. The overall expectation is that similar pairs of papers should have a similar number of incoming citations on average. We use a $z$-test to assess if that is the case (see Methods). This computation is performed in each PACS subfield separately to control for potential differences in the citation biases per subfield. 

Supplementary Table \ref{table:homa1} shows that in the majority of subfields, the average number of citations received by publications with male primary authors is higher than for female primary authors. In fact, we are able to observe a statistically significant difference in five out of ten subfields.

To check whether the temporal difference between two papers is responsible for the citation disparity for women (an older paper has had more time to accumulate citations), we add a maximum three-year difference restriction between two similar papers and redo the citation difference analyses. Table \ref{table:homa3} shows that when the time constraint is applied, the citation difference between two similar publications is no longer significant.

\begin{table}[h!]
    \centering
        \begin{tabular}{ccccccccc}
        \toprule
        PACS & Subfield & $N_{mf}$ & $p^*$ & $|M(p^*)|$ & $\frac{|M(p^*)|}{N_{mf}}$ & $d(p^*)$ & $z$ & \textit{p}-value \\ \midrule
        00 & General Physics & 32046 & 0.0009 & 1678 & 5.24\% & -0.443 & -0.582 & 0.561\\ 
        10 & Elementary Particles and Fields & 8502 & 0.0012 & 343 & 4.03\% & 0.650 & 0.843 & 0.399\\ 
        20 & Nuclear Physics & 1416 & 0.003 & 83 & 5.86\% & 0.494 & 0.697 & 0.486\\ 
        30 & Atomic and Molecular Physics & 5956 & 0.0012 & 255 & 4.28\% & 1.043 & 1.298 & 0.194\\ 
        40 & \begin{tabular}{@{}c@{}}Electromagnetism, Optics, Acoustics, Heat\\Transfer, Classical Mechanics, Fluid Dynamics\end{tabular} & 8180 & 0.0012 & 513 & 6.27\% & 0.583 & 0.919 & 0.358\\ 
        50 & Gases, Plasmas, Electric Discharges & 165 & 0.0038 & 6 & 3.64\% & 0.333 & 0.304 & 0.761\\ 
        60 & Condensed Matter (CM): Mechanical, Thermal & 8862 & 0.0006 & 485 & 5.47\% & 0.085 & 0.178 & 0.859\\ 
        70 & CM: Electrical, Magnetic, Optical & 41224 & 0.0005 & 2505 & 6.08\% & 0.505 & 1.265 & 0.206\\ 
        80 & Interdisciplinary Physics \& Related Studies & 5767 & 0.0015 & 265 & 4.60\% & 0.291 & 0.516 & 0.606\\ 
        90 & Geophysics, Astronomy, Astrophysics & 4651 & 0.003 & 248 & 5.33\% & 1.323 & 1.552 & 0.121 \\ \bottomrule
        \end{tabular}
        \caption{\textbf{Differences in received citations among similar pairs of publications with time restriction.} Gender differences in received citations among pairs of publications with validated similarity measured by $z$-scores. The variables of the columns are the following (more details in Methods): $N_{mf}$ - number of all possible male-female pairs; $p^*$ - chosen critical similarity value, the lower, the more similar; $M(p^*)$ - subset of pairs with similarity of $p^*$ or better; $d(p^*)$ - average male-female citation difference; $z$ - normalized difference of male-female average citations. Values of $p^*$ are chosen to establish $\frac{|M(p^*)|}{N_{mf}}$ values between 4\% and 7\%. Significant $z$-scores are marked in bold. In this table a maximum publication time difference of three years between the two papers of each validated pair has been considered.}
        \label{table:homa3}
    \end{table}




\subsection*{First-mover advantage drives the citation inequality}
Given the above results, we now seek to confirm whether the time of publication is a main driver for the citation disparity and whether the first-mover advantage in publication affects male-led papers and female-led papers similarly. We define $\Delta_t = Y_{m} - Y_{f}$ as the year difference between 
the publication dates of male-female pairs of similar papers and $\Delta_C = c_{m} - c_{f}$ as their citation difference. We plotted the year difference $\Delta_t$ against the citation difference $\Delta_C$ in Figure \ref{fig:centper}B. We likewise elaborated ten analogous plots after categorizing the data into subfields by PACS number (shown in Supplementary Figure \ref{fig:cyds}) to control for variations between subfields. Note that for this analysis we impose no time restriction between the publication times of the two papers of each pair.

To verify that the disparity in citations is caused by the first-mover advantage, we first need to test whether a first-mover advantage in fact exists. If that is the case, when a man publishes first ($\Delta_t<0$) he should get more citations ($\Delta_C>0$) on average, but when a woman publishes first ($\Delta_t>0$) she is the one who should get more citations ($\Delta_C<0$) on average; that is, in Figure \ref{fig:centper}B, quadrants Q2 and Q4 should be more populated than expected if we treated $\Delta_t$ and $\Delta_C$ as independent random variables. 
Equivalently, we should observe a negative correlation between $\Delta_t$ and $\Delta_C$. 


To test this hypothesis, we compared the empirical joint probability distribution of $\Delta_t$ and $\Delta_C$ ($P_\text{emp}(\Delta_t, \Delta_C)$) with the one that we would obtain if they were independent variables ($P_\text{null}(\Delta_t, \Delta_C)=p(\Delta_t)p(\Delta_C)$) by computing the \emph{probability anomaly} as:

\begin{equation}
    P_\text{diff}(\Delta_t,\Delta_C) = \frac{P_\text{emp}(\Delta_t, \Delta_C)-P_\text{null}(\Delta_t, \Delta_C)}{P_\text{null}(\Delta_t, \Delta_C)}
\label{eq:prob_anom}
\end{equation}

The resulting values of $P_\text{diff}(\Delta_t, \Delta_C)$ are shown in the right panel of Figure \ref{fig:centper}B and, as can be observed, they support the hypothesis of the first-mover advantage, since Q2 and Q4 present positive anomalies while Q1 and Q3 present negative ones. It is worth emphasizing that a positive (resp. negative) anomaly indicates higher (resp. lower) density of points with respect to a situation of no correlation between $\Delta_t$ and $\Delta_C$. To quantify this trend we computed the Pearson and Spearman correlations between $\Delta_t$ and $\Delta_C$, obtaining $-0.19$ and $-0.41$ respectively.

Once the existence of the first-mover advantage has been confirmed, we need to test whether there exists an \emph{asymmetry} in the relative advantage that men and women obtain when they publish first. If there is no asymmetry, the average number of citations that a woman obtains by publishing a certain number of years ahead of a man should be comparable to the number of citations that a man obtains in the equivalent situation. 

To verify this, we compared the citation differences of Q2 with Q4 (pairs where the earlier paper received more citations) and Q1 with Q3 (pairs where the earlier paper received fewer citations) for each temporal difference; in other words, we compared the average absolute value $\left|\Delta_C\right|$ of points from Q2 with the average $\left|\Delta_C\right|$ of points from Q4 for each $\left|\Delta_t\right|=1,2,\dots$ separately (analogously for Q1 and Q3). To perform this comparisons, we used $z$-tests for difference of means for each year difference (see Methods). The results of the tests for every subfield, shown in Supplementary Table \ref{table:cyd}, indicate that there is no significant disparity in the advantage obtained by women and men when they publish a paper a given number of years earlier than their corresponding statistical twin.

This thorough analysis indicates that, when we control for the similarity of the papers and time of publication, there is no significant evidence for any disparity between two statistical twins. Therefore, despite the common assumption that papers written by women generally receive fewer citations, this difference is mainly driven by the historical first-mover advantages that men have and not by deliberate discriminatory actions against women.

\subsection*{Historical trend in citation}
Finally, we hypothesize that the physics community might have been less receptive to the contribution of women in the past compared to the present. To test this hypothesis, we measure the temporal evolution of the centrality differences ($\Delta_C$) between male-female pairs by year and limit the publication time difference between the two papers to a trailing window of three years. Then, we compute the mean and standard error of $\Delta_C$ for all the pairs within each window. 
For comparison, we perform an analogous computation for random samples of similar male-male pairs. In each time window, we matched the number of sampled male-male pairs with the number of similar male-female pairs. We repeated the male-male computation 100 times 
independently and computed the average $\Delta_C$ and the standard error, which we use as a baseline. 


Figure \ref{fig:centper}C shows the citation differences for male-female pairs of similar papers over the years compared with the baseline given by male-male pairs of papers. The earlier male-female pairs seem to present a higher disparity favoring men than later pairs, whereas the $\Delta_C$ values for male-male pairs throughout the years are, as expected, consistently located around the equilibrium. After all, the similar male-male pairs were chosen randomly and there is no reason for one paper of the pair to have a higher or lower citation count than the other. The number of sampled pairs per year is shown in Supplementary Figure \ref{fig:simcount}.
To measure the apparent change of trend in the male-female pairs, we ran a Mann-Whitney U Test comparing the $\Delta_C$ of male-female pairs published before and after 1995, obtaining a \textit{p}-value$=6.9 \times 10^{-9}$. Hence, as hypothesized, the male-female pairs published before 1995 show a significant disparity favouring men when compared to those published after 1995.

\section*{Discussion}

The primary objective of this research was to identify gender disparities in physics focusing on five topics of interest: productivity, author order analysis, self-citation analysis, and the comparison of citations for pairs of similar papers. Therefore, our study makes a substantial contribution to the current body of literature by comprehensively analysing the citation patterns of men and women in physics. We assembled information about all papers published in the American Physical Society from 1894 to 2009. Using a technique that combines name and image recognition, we inferred the gender of the primary authors of papers and, to study potential gender biases, we looked for statistically significant differences in the citation patterns of papers written by male and female primary authors.

Despite all the efforts to avoid any biases in our analysis, some caveats should be considered. We have combined name and image inference to identify the gender of the scholars. Even with this careful examination, we cannot infer the gender of authors who have only initials as their first names. 
Another caveat is related to ethnicity, as we cannot identify the majority of Asian names originating from Korea and Japan \cite{karimi2016inferring}. However, we can safely argue that this lack of gender identification likely affects both genders similarly. Another sensitive step of our data processing pipeline is name disambiguation, used to identify all the papers published by a given author. Although we have used various criteria to disambiguate names, there still might be errors in identifying unique authors and these errors may affect minorities, which have lower numbers of instances in the data. There are other factors that can affect citation and may not be determined by assessing similar papers. For example, papers that are novel and ground-breaking or interdisciplinary in their nature may contain citations from outside physics that make them less similar to other established papers, and those are likely not being assessed in our analysis. In this case, we acknowledge that the focus of our analysis is on those scholars who work predominantly on mainstream physics. 






\subsection*{Broader impact}

\paragraph{Academic evaluation metrics.}
The academic community tends to evaluate scientists based on the behaviour of the majority, which in physics is predominantly the behaviour of white, Western men. This evaluation, at its core, is problematic and can cause discrimination against other groups that are historically, socially, or politically discriminated against. In such cases, more attention and care should be given to women and other minorities who are more likely to suffer from such historical disadvantages. Once the system moves towards a more diverse representation, its core values will no longer be determined by only one type of majority. 

\paragraph{Structural inequalities and cumulative effects.}
The structure of the citation network can influence the future citations and recognition that papers receive. Through reading papers, scholars often follow cited papers to read and cite previous works. If papers written by women are under-represented in influential positions of the citation network, this will affect their future visibility even if they are cited adequately compared to their statistical twins. This phenomenon, also known as success-breeds-success \cite{van2014field}, in addition to cumulative advantages and the first-mover advantage \cite{newman2009first}, can be consequential for the success and recognition of scholars, their visibility \cite{karimi2018homophily}, future success, and the scientific community's perception of their work \cite{lee2019homophily}. 

\paragraph{Collaboration barriers.}
Science, at its core, is a collaborative process. Through collaboration and research visits, scientists meet, ideas spread, and the foundations are laid for future collaborations. There are implicit factors that can indirectly affect the participation of women in scientific collaborations. For example, geographical distance is more likely to affect women due to their family responsibilities, restrictions on travel during pregnancy, and breastfeeding, to name a few reasons. Women might not be welcomed in certain social events that are predominantly preferred by men or for those with no family responsibilities. Lack of chemistry or shyness in interacting with another gender might also make women less likely to be invited for research visits and collaborations. We note that women are not the only group who suffer from geographical restrictions, as other forms of discrimination or simply high traveling costs can affect the collaboration of scholars from Muslim and developing countries.

\paragraph{The importance of diversity.}
Diversity has a crucial role in shaping and spreading new ideas.  For example, one can safely argue that many recent publications that aim to understand the inequality and biases in academia and other social domains are directly related to the boost in participation of women and minorities. However, it is also known that despite their contributions to innovative research, minorities do not reap the benefits of their innovation when compared with majorities \cite{hofstra2020diversity}. In future work, intersectional inequalities should be studied at large scale by considering the intersection of gender, ethnicity, and race. 






\section*{Conclusion}
In sum, we found that despite the rise of female participation in physics in recent years, the rate of entry of new women into the field is still much slower than for men. Women tend to be less productive than men in their mid-career, and they tend to have a higher dropout rate over their academic careers. Moreover, in agreement with previous works, we found that men tend to cite their own previous works with more frequency than women, penalizing the visibility of women and their potential for academic promotion. This disparity in visibility is also manifested in the under-representation of women at the top ranks of both degree and PageRank centrality of the citation network, which implies a disadvantage on both a local scale (lower number of citations) and a global scale (peripheral location within the network).

When assessing pairs of similar papers, we found that while earlier papers tend to have an advantage, there are no statistically significant differences in citations of men and women. These results combined suggest that the overall disparity in the citation network is a result of cumulative advantages and the first-mover effect that men have in physics, and not an intentional discriminatory act against women. This cumulative advantage, however, could create implicit biases that should be tackled by appropriate policies that foster the participation of women and other minorities.

\section*{Methods}

\subsection*{Assessing similar pairs of papers}

The main objective of this paper is to compare pairs of similar papers in an unbiased fashion. The similarity analysis
is based on the concept of bibliographic coupling strength $N_{ij}$ of pairs of articles $(i,j)$, which is defined as the number of common articles cited by both $i$ and $j$ \cite{kessler1963bibliographic, egghe2002co}.
To overcome the shortcomings of the most commonly used normalized versions of $N_{ij}$ (the Jaccard index and fractional counting, described in Supplementary Section 3), we identify couples of similar papers by looking both at the outgoing references of the pair and the incoming citations of the articles they cite. In particular, we perform a statistical test using the hypergeometric distribution as a null model and detect pairs of papers whose set of common outgoing citations has a very low probability of having been generated by chance \cite{tumminello2011statistically}. In Supplementary Figure \ref{fig:similarity_alg} we present a diagram of this methodology, which is explained below in detail.


First, the citation network is built for each physics subfield (the first two digits of PACS), and then each paper in the citation network is further labeled by the gender of its primary author. After establishing the citation network, two sets $S_A^k$ and $S_B^k$ are defined: 
$S_B^k$ includes all articles that are cited $k$ times, and $S_A^k$ includes all articles that cite any element in $S_B^k$. Notice that each publication may belong to one set, to the other or to both.

Then, 
we build all possible pairs $i, j \in S_A^k$. In order to quantify the similarity between $i$ and $j$, we compute the probability of $i$ and $j$ both referencing a certain number of publications using the hypergeometric distribution:

\begin{equation}
    H(X|N_B^k, d_i, d_j) = \frac{\binom{d_i}{X}\binom{N_B^k-d_i}{d_j-X}}{\binom{N_B^k}{d_j}}
    \label{eq:hypergeom}
\end{equation}

where $N_B^k = |S_B^k|$ and $d_i$, $d_j$ are the number of elements in $S_B^k$ that publications $i$ and $j$ respectively cite. Supplementary Figure \ref{fig:similarity_alg} shows a diagram that illustrates the meaning of these variables. Notice that if $d_i$ and $d_j$ are interchanged, the value of $H$ remains the same. Finally, $X$ would be the number of overlapping citations. The term $\binom{N_B^k}{d_j}$ corresponds to all the possible ways of choosing $d_j$ publications from the set $S_B^k$; $\binom{d_i}{X}$ denote the number of ways one can choose exactly $X$ publications from the $d_i$ papers that $i$ cites and $\binom{N_B^k-d_i}{d_j-X}$ are the number of ways the $d_j-X$ papers cited by $j$ and not by $i$ can be chosen from $S_B^k$. Intuitively, this hypergeometric distribution can be understood as an urn model with $N_B^k$ balls, such that $d_i$ of them are \emph{good} balls and the rest are \emph{bad} balls. $H$ is then the probability of obtaining exactly $X$ \emph{good} balls when retrieving $d_j$ balls from this urn.

Now, if $i$ and $j$ have actually cited $N^k_{ij}$ common papers of in-degree $k$, the cumulative probability of $X\leq N^k_{ij}$ provides a measure of how probable it is that the size of their set of overlapping citations can be explained by randomness:

\begin{equation}
    p_{ij}(k) = \sum_{X=0}^{N_{ij}^k-1}{H(X|N_B^k, d_i, d_j)}
    \label{eq:pij}
\end{equation}



The higher $p_{ij}(k)$ is, the less probable it is that the size of $N^k_{ij}$ is due to chance. Therefore, we devise a measure of \textit{similarity} as follows:

\begin{equation}
    q_{ij}(k) = 1 - p_{ij}(k)
\end{equation}

Notice that $q_{ij}(k)$ is the probability of a particular bibliographic coupling strength of randomly selected papers $i$ and $j$ towards articles in $S_B^k$ being greater than or equal to $N_{ij}^k$. This computation is repeated for all $k$ and the different values of $q_{ij}(k)$ are stored. The similarity of the couple $(i,j)$ is measured by the minimum overall possible values of $k$:

\begin{equation}
    {q_{ij}(k)}_{\min} = \min_{k} \{1 - q_{ij}(k) \}
\end{equation}

Publications $i$ and $j$ are considered similar if ${q_{ij}(k)}_\text{min} < p^*$, where $p^*$ is a threshold value. By studying ${q_{ij}(k)}_\text{min}$, we are now able to assess and compare pairs of similar papers. 


In addition, we manually inspected several pairs of papers with validated similarity measurements to verify the accuracy of our approach. We set a low threshold value, $p^* = 10^{-6}$, and applied a constraint of maximum publication year difference of three years. We validated the similarity between the two papers through the inspection of keywords, titles, and citation activities. For instance, papers \cite{tannenbaum2005evolutionary} and \cite{lee2007asexual}, with ${q_{ij}(k)}_{\min} = 6.1969 \times 10^{-7}$, present some connection between their main ideas and share a common author. Additionally, a large proportion of their citation activities align. Another similar pair is formed by articles \cite{reimann2002electronic} and \cite{szafran2005three} with ${q_{ij}(k)}_{\min} = 5.0855 \times 10^{-8}$, which show extremely similar citation activities and deal with similar topics. As a final example, \cite{casademunt1989decay} and \cite{ramrez1991first}, with ${q_{ij}(k)}_{\min} = 2.8198 \times 10^{-8}$, share topic, citation activities, and a collaborating author. It is worth emphasizing that, due to the highly restrictive $p^*$, some of these statistically validated pairs of similar papers share a common author, which is a strong verification of our algorithm.

In a nutshell, the hypergeometric probability testing compares how significant the overlapping outflow of citations is for two papers compared to what we expect from the in-degree and out-degree of the citation network. Using this technique, we are able to compare papers that are inherently similar in their subject field by not only comparing their overlapping references, but also accounting for variations in the citations received by each reference. Since we control both for the outgoing citations of the pair and the incoming citations of the commonly cited papers, the comparison is robust and unbiased.
\subsection*{Authorship order two-proportion $z$-test}
We denote the total male and female population as $N_{m}$ and $N_{f}$, and total number of male and female first authors as $n_{m}$ and $n_{f}$, respectively. We further define $p_{m} = \frac{n_{m}}{N_{m}}, p_{f} = \frac{n_{f}}{N_{f}},p = \frac{n_{m}+n_{f}}{N_{m}+N_{f}}$ and the two-proportion $z$-test is performed as below:

\begin{equation}
    z = \frac{p_{m}-p_{f}}{\sqrt{p\left(1-p\right)\left(\frac{1}{N_{m}} + \frac{1}{N_{f}}\right)}}
    \label{eq:z_2}
\end{equation}

\subsection*{Calculating differences in received citations}
\label{sec:diff_cit}


Let $N_{mf}$ denote the cardinality of the set of all pairs $(m,f)$ where $m$ and $f$ denote publications by a primary male and female author respectively and let $M(p^*)$ be the subset of all similar pairs validated under $p^*$. $c_m$ and $c_f$ indicate number of citations received by $m$ and $f$, and the difference in number of received citations $c_d$ can be computed by

\begin{equation}
c_d(p^*) = \sum_{x=1}^{|M(p^*)|}{(c_m-c_f)_x}
\label{eq:cen}
\end{equation}

\noindent where $x$ denotes the index of pairs $(m,f) \in M(p^*)$. We define an average centrality difference per subfield as

\begin{equation}
    d(p^*) = \frac{c_d(p^*)}{|M(p^*)|}
    \label{eq:cenavg}
\end{equation}

\noindent We perform a difference of means $z$-test with $H_0: c_m = c_f$, with the $z$-statistic defined as

\begin{equation}
    z = \frac{\bar{c}_m-\bar{c}_f}{\sqrt{\frac{\sigma^2_{c_m}}{|M(p^*)|}+\frac{\sigma^2_{c_f}}{|M(p^*)|}}}
    \label{eq:z}
\end{equation}

\noindent Hence, a positive $z$-score indicates that the data displays higher degree centrality for male authors than expected.




\subsection*{Computing temporal citation differences}

We compared the citation differences of Q2 with Q4 (pairs where the earlier paper received more citations) and Q1 with Q3 (pairs where the earlier paper received fewer citations) for each temporal difference; in other words, we compared the average absolute value $\left|\Delta_C\right|$ of points from Q2 with the average $\left|\Delta_C\right|$ of points from Q4 for each $\left|\Delta_t\right|=1,2,\dots$ separately (analogously for Q1 and Q3). To perform these comparisons, we used $z$-tests for difference of means for each year difference:

\begin{equation}
   z = \frac{\overline{|\Delta_C^{Q_i}|}-\overline{|\Delta_C^{Q_j}}|}{\sqrt{\frac{\sigma^2_{Q_i}}{N(Q_i)}+\frac{\sigma^2_{Q_j}}{N(Q_j)}}}
    \label{eq:zcit}
\end{equation}

In this test we evaluate the mean ($\overline{|\Delta_C^{Q_i}|}$) and the standard deviation ($\sigma_{Q_i}$) of $\left|\Delta_C\right|$ for two subsets of quadrants $Q_i$ and $Q_j$. $N(Q_i)$ is the number of data points in quadrant $i$ (number of similar pairs). We run the $z$-test for $(i, j) = (1, 3)$ and $(i, j) = (2, 4)$. 


\bibliography{nature}



\section*{Acknowledgments}

This work has been supported by the Austrian research agency (FFG) under project No. 873927. We would also like to thank M. R. Ferreira, J. Bachmann, G. Amichay, and S. Sajjadi for their comments and suggestions, which helped to greatly improve the manuscript. And to J. Reddish, for her incredibly thorough proofreading of the paper.

\clearpage

\section*{Supplementary Information}

\section{Author name disambiguation}

\begin{figure}[h!]
         \centering
         \includegraphics[width=\textwidth]{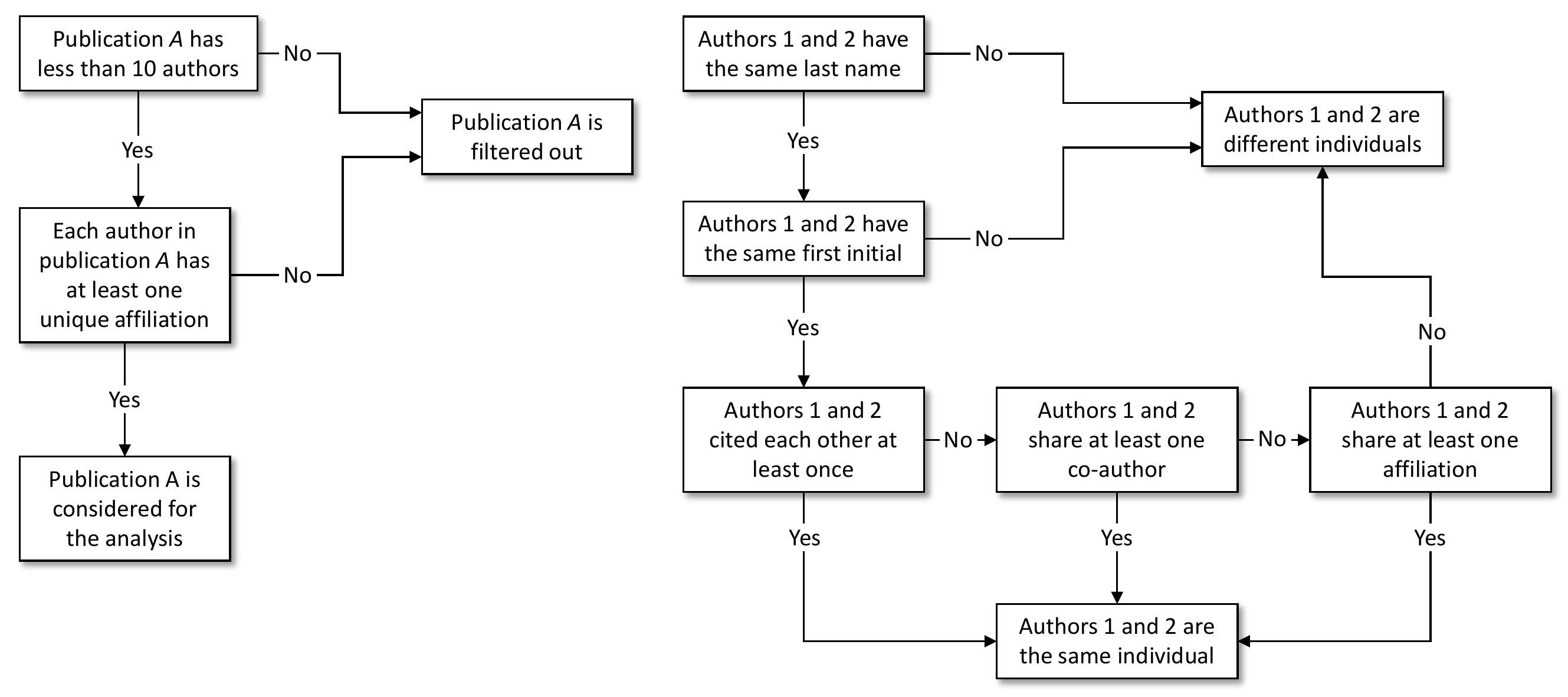}
         \caption{\textbf{Author name disambiguation algorithm.} This flow chart schematizes the author name disambiguation algorithm that Sinatra \textit{et al}. used \cite{sinatra2016quantifying}. The algorithm first decides whether a publication is considered in the analysis. Then, for any two author names 1 and 2, it decides whether they are the same individual or two different authors.}
     \label{fig:disamb}
     \end{figure}

We used a preprocessed version of the APS dataset based on Sinatra et al. \cite{sinatra2016quantifying} where a name disambiguation method was applied to correctly match every author to her papers. With this method, summarized in the flow chart of Figure \ref{fig:disamb}, 237,000 different authors were identified.

\section{Gender detection}

In order to detect the gender from authors' names, the first step is to remove those authors whose first name is not mentioned and initialized. No existing name-based gender inference techniques can tackle those cases. For those authors that we had first names available, we first use the application \textit{Genderize} \cite{genderize}. Then, for the names whose gender this application is unable to infer, we use the picture-based gender inference technique \textit{Face++} \cite{zhou2015naivedeep}. In this second step, we perform a Google image search with the author's first name and family name, and feed the resulting images to \textit{Face++}. This methodology was developed by Karimi \emph{et al.} \cite{karimi2016inferring}, who compared it with commonly used dictionary-based gender detection techniques and showed that it consistently achieves high accuracy for names of different nationalities. The results they obtained for a random sample of researchers whose names and genders are known are shown in Table \ref{table:4}. 

As a preliminary step to use the gender detection technique we performed a thorough standardization of names to avoid issues with the use of special characters. We followed the rules from the Program for Cooperative Cataloguing of the Library of Congress (NACO) \cite{naco}. Supplementing the NACO normalization by translating accented characters and other special characters accordingly improves the overall query matching by 63\%. Using this methodology we were able to detect the gender of 124,000 authors.

\begin{table}[]
\centering
\begin{tabular}{cccccccc}
\toprule
 & Sample Size & \textit{SSA} & \textit{IPUMS} & \textit{Sexmachine} & \textit{Genderize} & \textit{Face++} & \begin{tabular}[c]{@{}c@{}}\textit{Genderize}\\ \& \textit{Face++}\end{tabular} \\ \midrule
United States & 419 & 82\% & 76\% & 84\% & 83\% & 91\% & 91\% \\
China & 113 & 20\% & 11\% & 67\% & 28\% & 65\% & 50\% \\
United Kingdom & 96 & 94\% & 92\% & 92\% & 94\% & 81\% & 98\% \\
Germany & 82 & 87\% & 88\% & 96\% & 94\% & 87\% & 96\% \\
Italy & 75 & 93\% & 92\% & 94\% & 98\% & 79\% & 99\%  \\
Canada & 60 & 87\% & 77\% & 86\% & 91\% & 90\% & 96\% \\
France & 58 & 93\% & 92\% & 80\% & 96\% & 81\% & 97\% \\
Japan & 56 & 79\% & 70\% & 100\% & 90\% & 62\% & 91\% \\
Brazil & 44 & 29\% & 29\% & 15\% & 44\% & 81\% & 90\% \\
Spain & 39 & 96\% & 92\% & 92\% & 100\% & 92\% & 100\%  \\
Australia & 31 & 89\% & 89\% & 90\% & 86\% & 86\% & 94\% \\
India & 29 & 67\% & 17\% & 71\% & 78\% & 83\% & 83\% \\
South Korea & 27 & 4\% & 0\% & 58\% & 11\% & 74\% & 37\% \\
Switzerland & 25 & 78\% & 70\% & 56\% & 83\% & 88\% & 90\% \\
Turkey & 21 & 43\% & 14\% & 79\% & 81\% & 86\% & 100\% \\\bottomrule
\end{tabular}
\caption{\textbf{Comparison of gender detection techniques.} This table compares the accuracy of various gender inference methodologies for names from 16 different countries. The dictionary-based methods are respectively based on the US Social Security Administration (SSA), the survey from Integrated Public Use Microdata Series (IPUMS), and the publicly available list of names \textit{Sexmachine}. This table is a modified version of Table 2 of \cite{karimi2016inferring}.
}
\label{table:4}
\end{table}

\section{Similarity measures for publications}


The main objective of this paper is to compare pairs of similar papers in an unbiased fashion. The similarity analysis
is based on the concept of bibliographic coupling strength $N_{ij}$ of pairs of articles $(i,j)$, which is defined as the number of common articles cited by both $i$ and $j$ \cite{kessler1963bibliographic, egghe2002co}. But using $N_{ij}$ without further considerations can lead to misleading results. For example, the similarity between two papers that include each 20 and 25 citations and share $N_{ij}=5$ of them should not be the same as the similarity between two papers that also share 5 references but respectively cite 65 and 82 publications. On the other hand, within subfields there are usually a handful of very popular publications that are cited in most works (such as review papers), so their inclusion in two different papers may not indicate actual similarity. In order to obtain meaningful measures of similarity, several normalization approaches have been explored. 

A widely used measure that addresses the first kind of the issues described above is the Jaccard index. The Jaccard index is computed as the quotient of the cardinality of the intersection and the cardinality of the union of the sets of cited publications by the two papers under consideration. One of the problems of this method is that it considers the weight of all citations to be identical and therefore does not take the significance of each paper into account \cite{lee2017improving, saranya2016performance}. In addition, narrowing our analysis to counting the common articles may not lead to an accurate interpretation due to the massive differences between male and female sample sizes. The reason is that, if the sizes of the sets of citations of the two papers are very different, their similarity is primarily determined by the size of the smallest one, as their intersection is bounded by the size of the smallest set \cite{ciotti2016homophily}.

\emph{Fractional counting} is another common normalization technique for bibliographic coupling. In this case, instead of normalizing by the outgoing citations of the two papers of interest, each commonly cited reference contributes to the similarity score with a weight inversely proportional to its number of incoming citations \cite{PERIANESRODRIGUEZ20161178,Batagelj2020}. Therefore, fractional counting addresses the second kind of situation discussed above by compensating the disproportionate influence of very popular publications in the similarity score. 
However, unlike the Jaccard index, it does not take into account the relative size of the sets of outgoing citations.

\begin{figure}[h]
    \centering
    \includegraphics[width=0.8\textwidth]{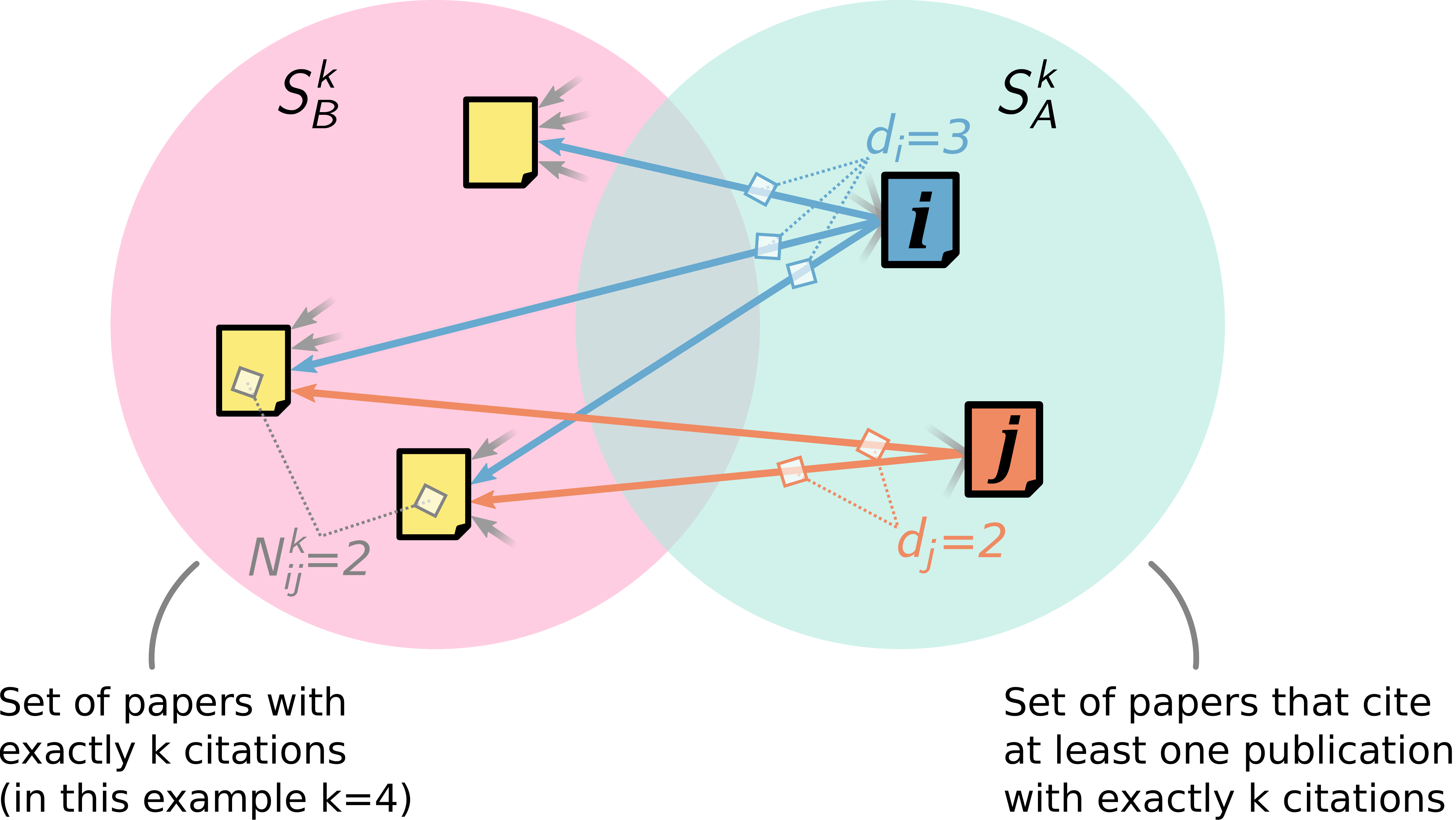}
    \caption{\textbf{Setting up for the similarity algorithm.} This figure sketches the variables involved in the computation of the hypergeometric distribution function used to obtain the paper similarity measure \cite{ciotti2016homophily}.}
\label{fig:similarity_alg}
\end{figure}

To overcome the issues of the Jaccard index and fractional counting, we identify couples of similar papers by looking both at the outgoing references of the pair and the incoming citations of the articles they cite. In particular, we perform a statistical test using the hypergeometric distribution as a null model and detect pairs of papers whose set of common outgoing citations has a very low probability of having been generated by chance \cite{tumminello2011statistically}. In Figure \ref{fig:similarity_alg} we present a diagram of this methodology, which is explained in Methods in detail.

\clearpage

\section{Additional supplementary figures}


\begin{figure}[h]
        \centering
        \includegraphics[width=0.5\textwidth]{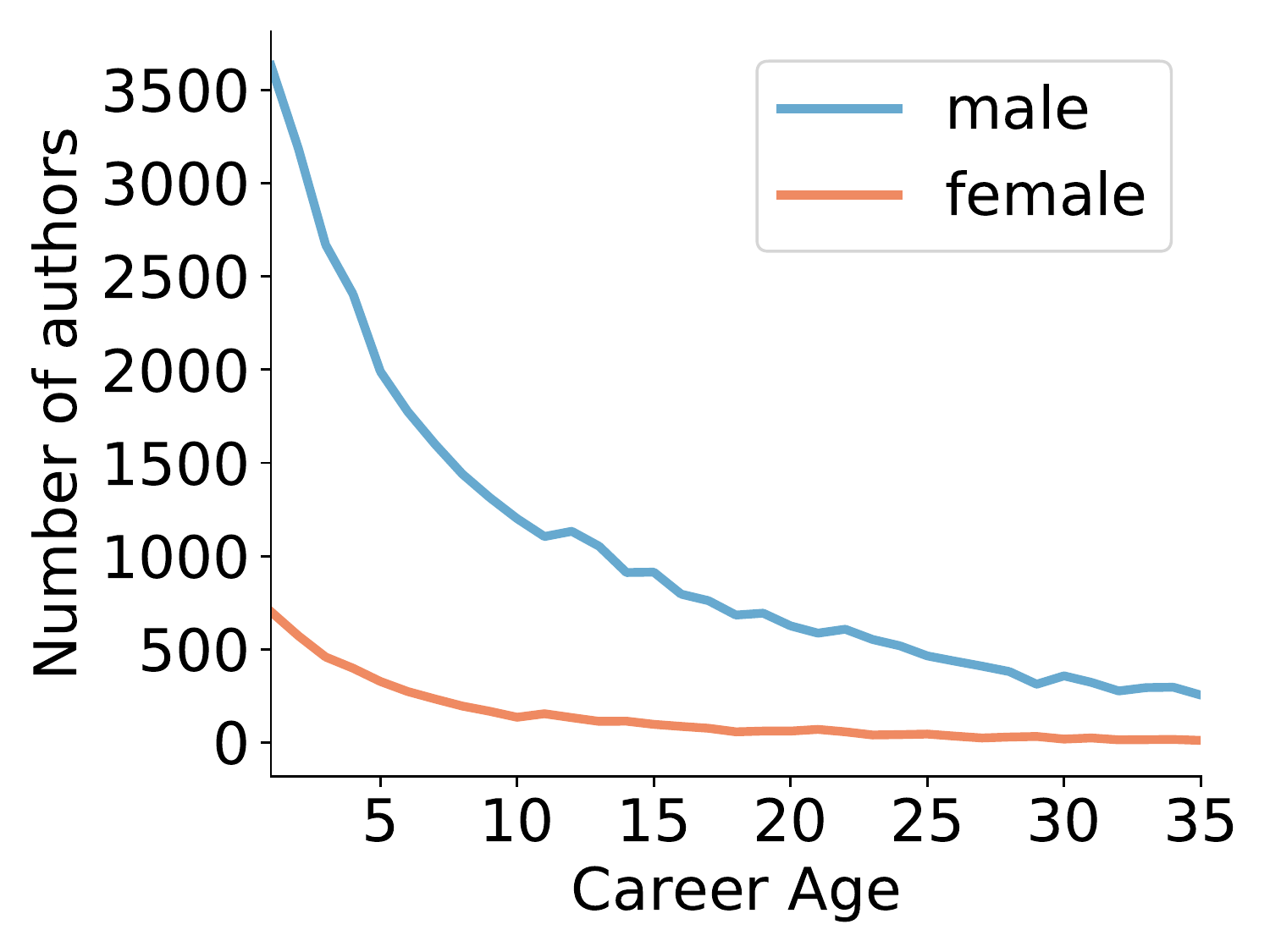}
        \caption{\textbf{Number of authors by career age.} Number of male and female authors by their career age. As mentioned in the main part of the paper, authors with career age of 0 were exempt from this analysis.}
        \label{fig:cacount}
    \end{figure}
    
\begin{figure}[h]
        \centering
        \includegraphics[width=\textwidth]{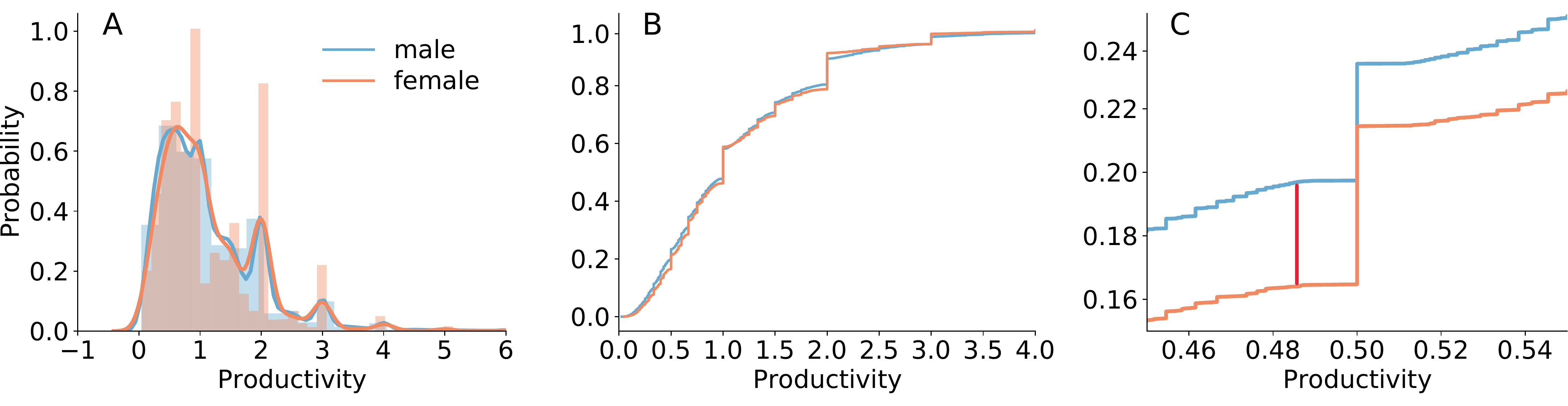}
        \caption{\textbf{Productivity distribution of authors by gender.} A: Probability density functions (PDF) of the productivity distributions of male and female APS authors. B: Corresponding empirical cumulative distribution functions (CDF) of the productivity distributions. C: Same as B but focused on the productivity interval with the maximum difference between the two distributions, indicated with a vertical red line.}
        \label{fig:cdfs}
    \end{figure}

\begin{figure}
\begin{subfigure}{.5\textwidth}
  \centering
  \includegraphics[width=.65\linewidth]{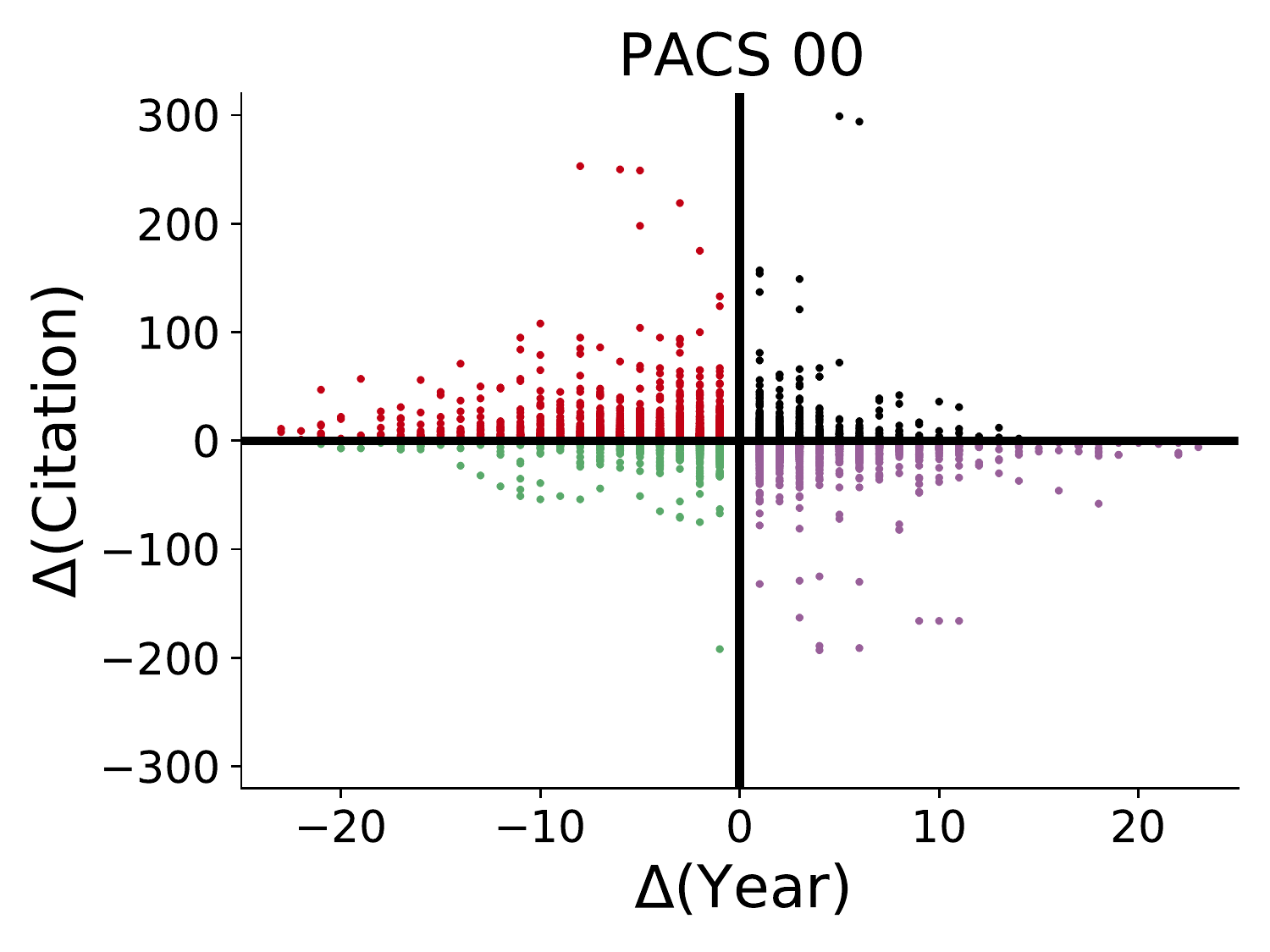}
\end{subfigure}%
\begin{subfigure}{.5\textwidth}
  \centering
  \includegraphics[width=.65\linewidth]{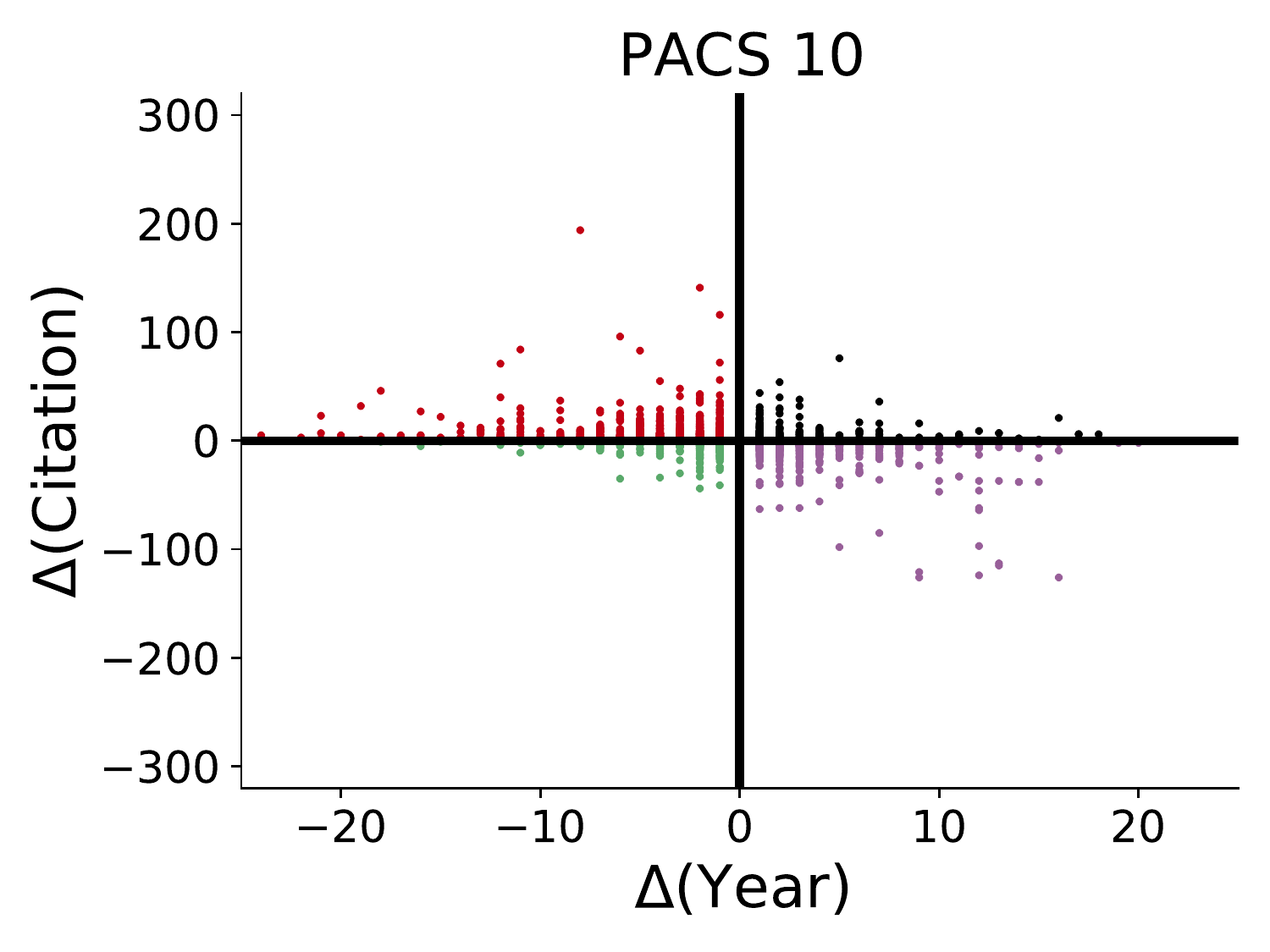}
\end{subfigure}
\begin{subfigure}{.5\textwidth}
  \centering
  \includegraphics[width=.65\linewidth]{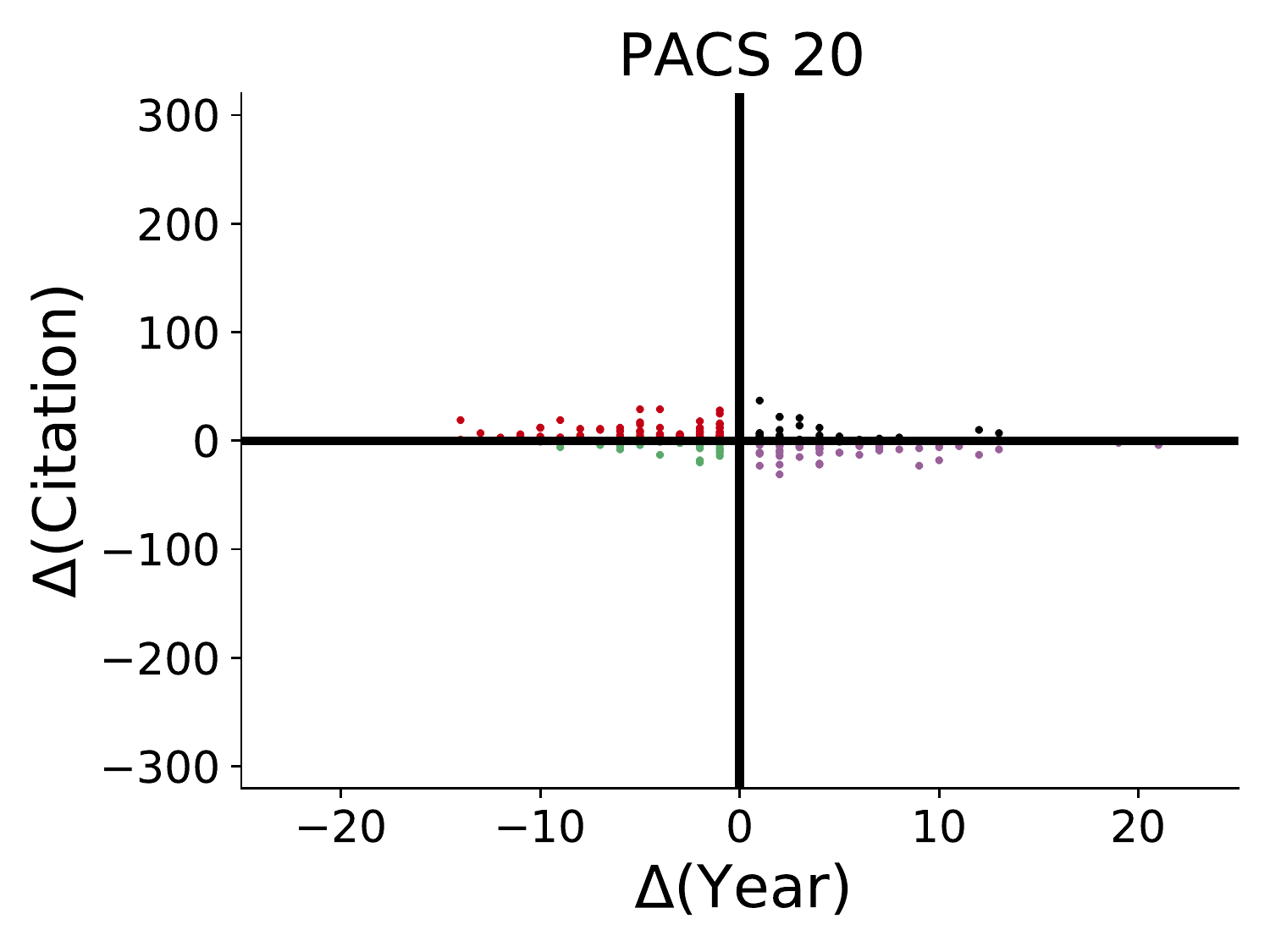}
\end{subfigure}%
\begin{subfigure}{.5\textwidth}
  \centering
  \includegraphics[width=.65\linewidth]{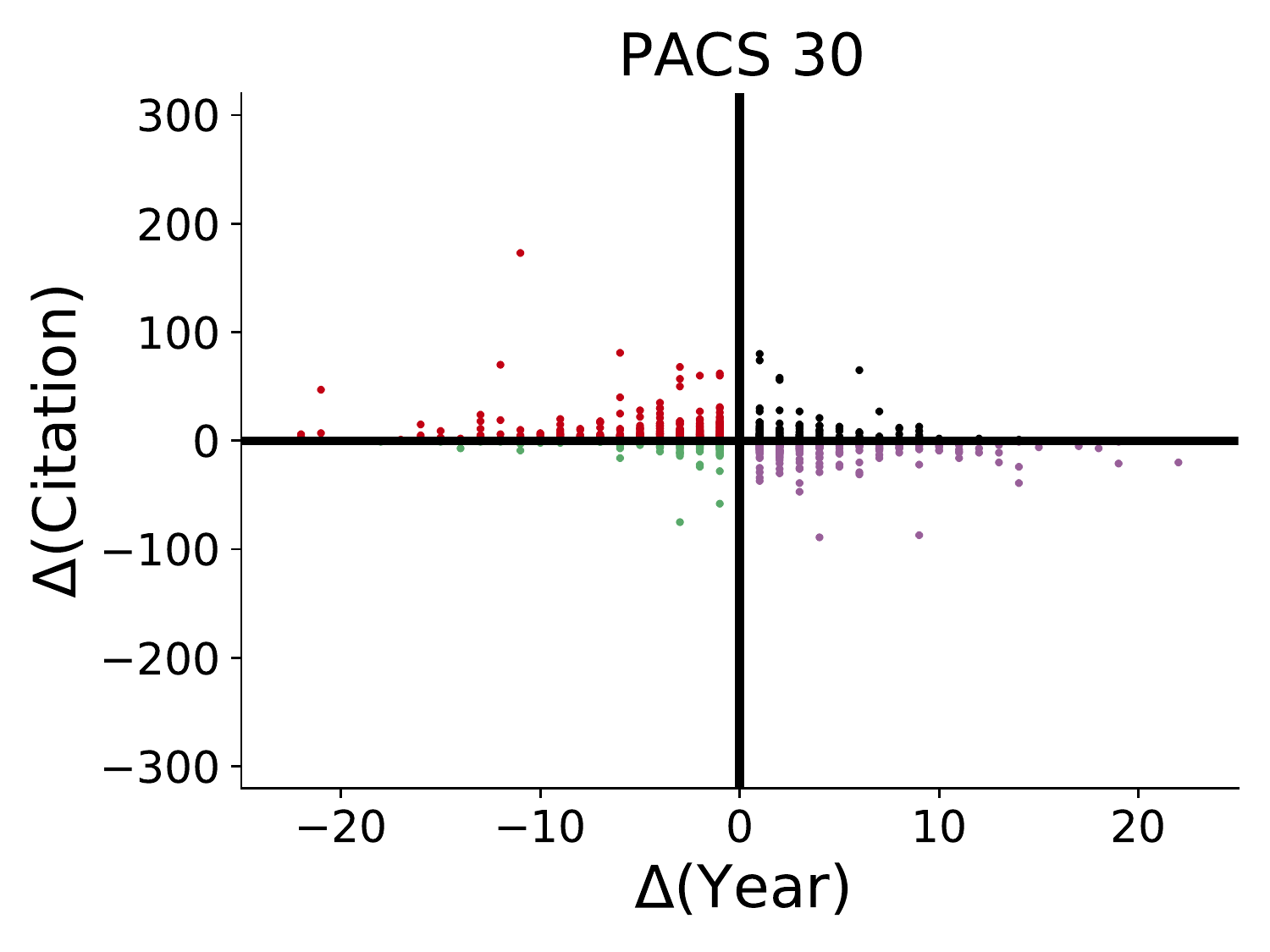}
\end{subfigure}
\begin{subfigure}{.5\textwidth}
  \centering
  \includegraphics[width=.65\linewidth]{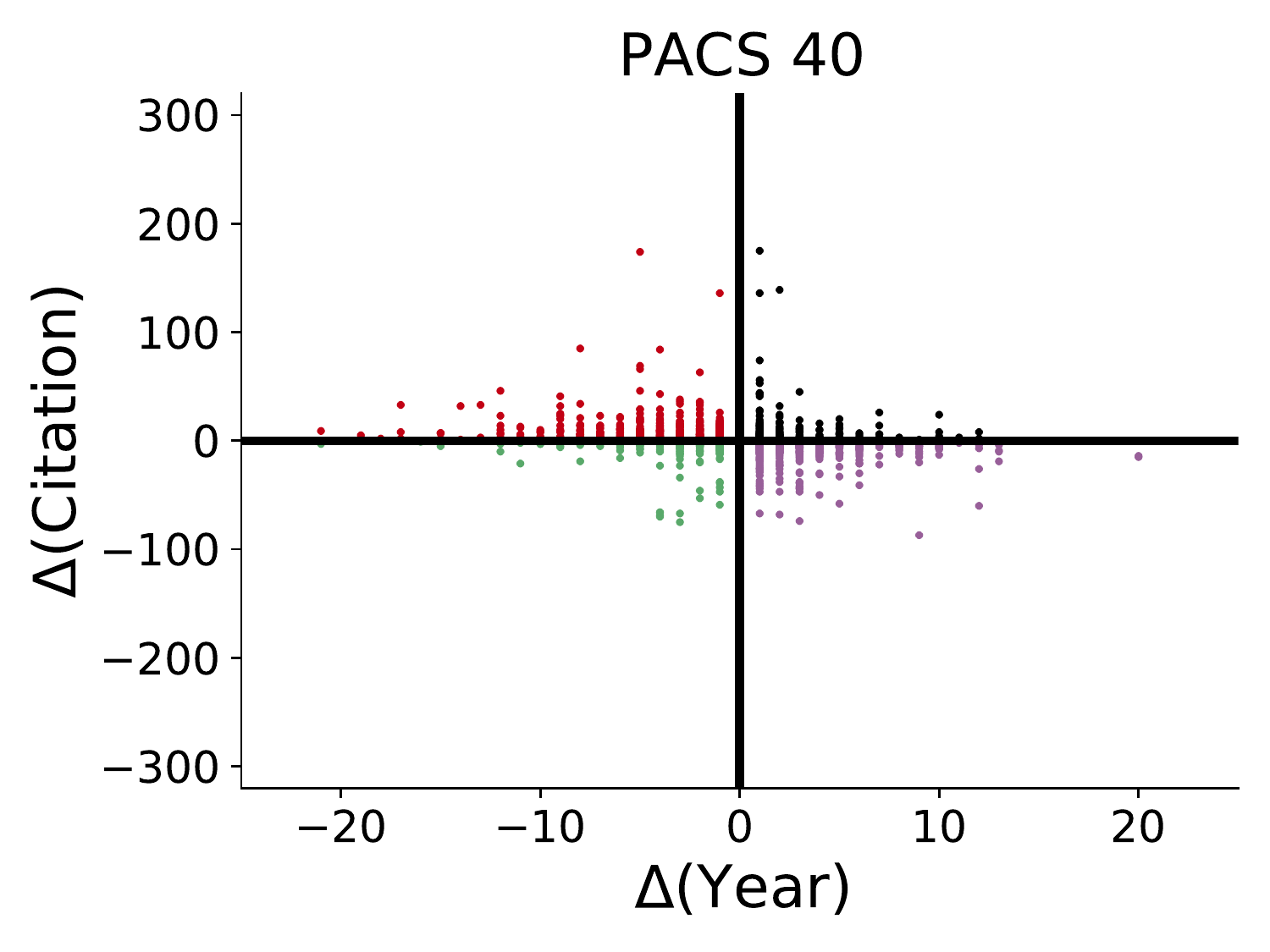}
\end{subfigure}%
\begin{subfigure}{.5\textwidth}
  \centering
  \includegraphics[width=.65\linewidth]{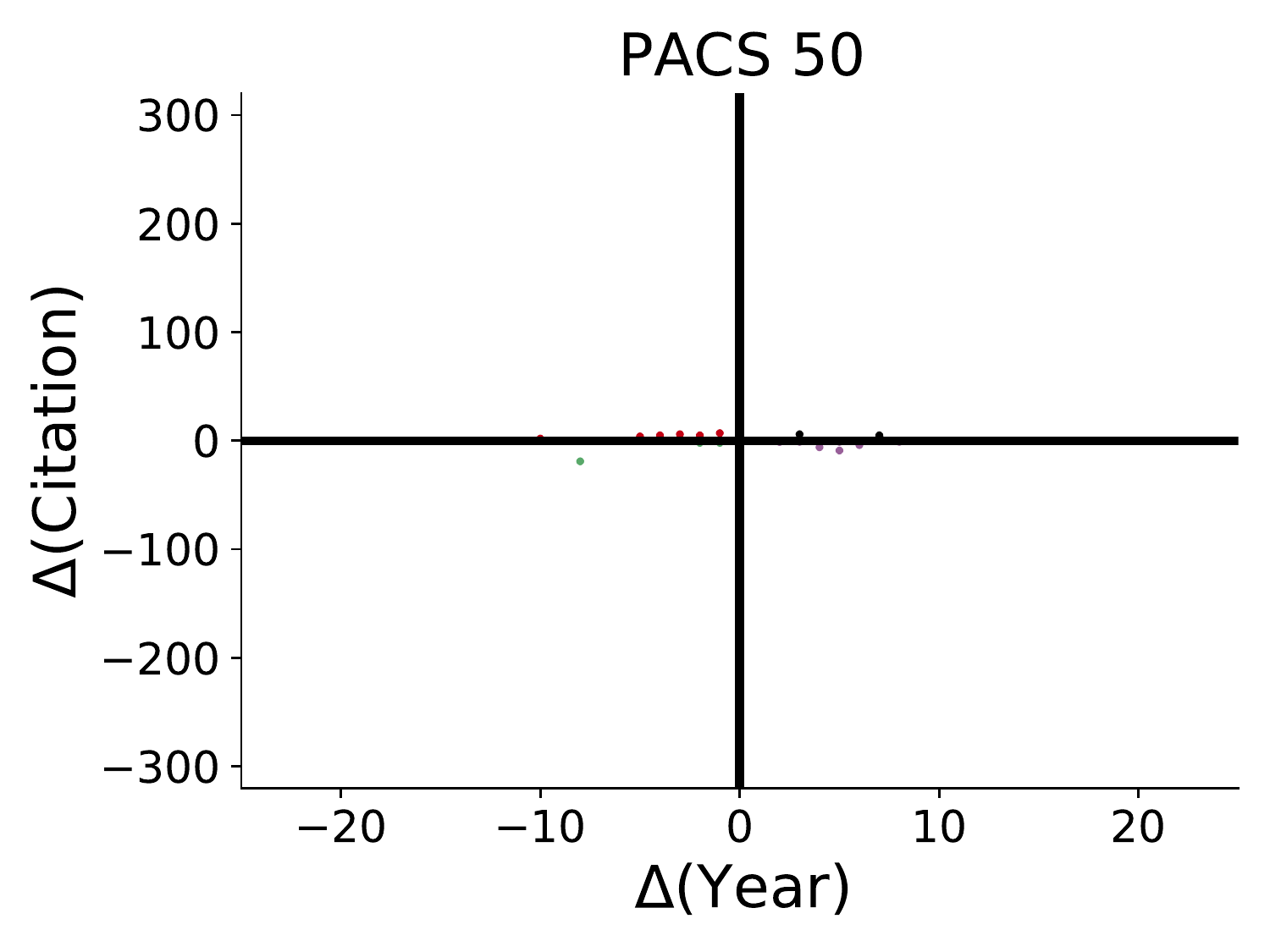}
\end{subfigure}
\begin{subfigure}{.5\textwidth}
  \centering
  \includegraphics[width=.65\linewidth]{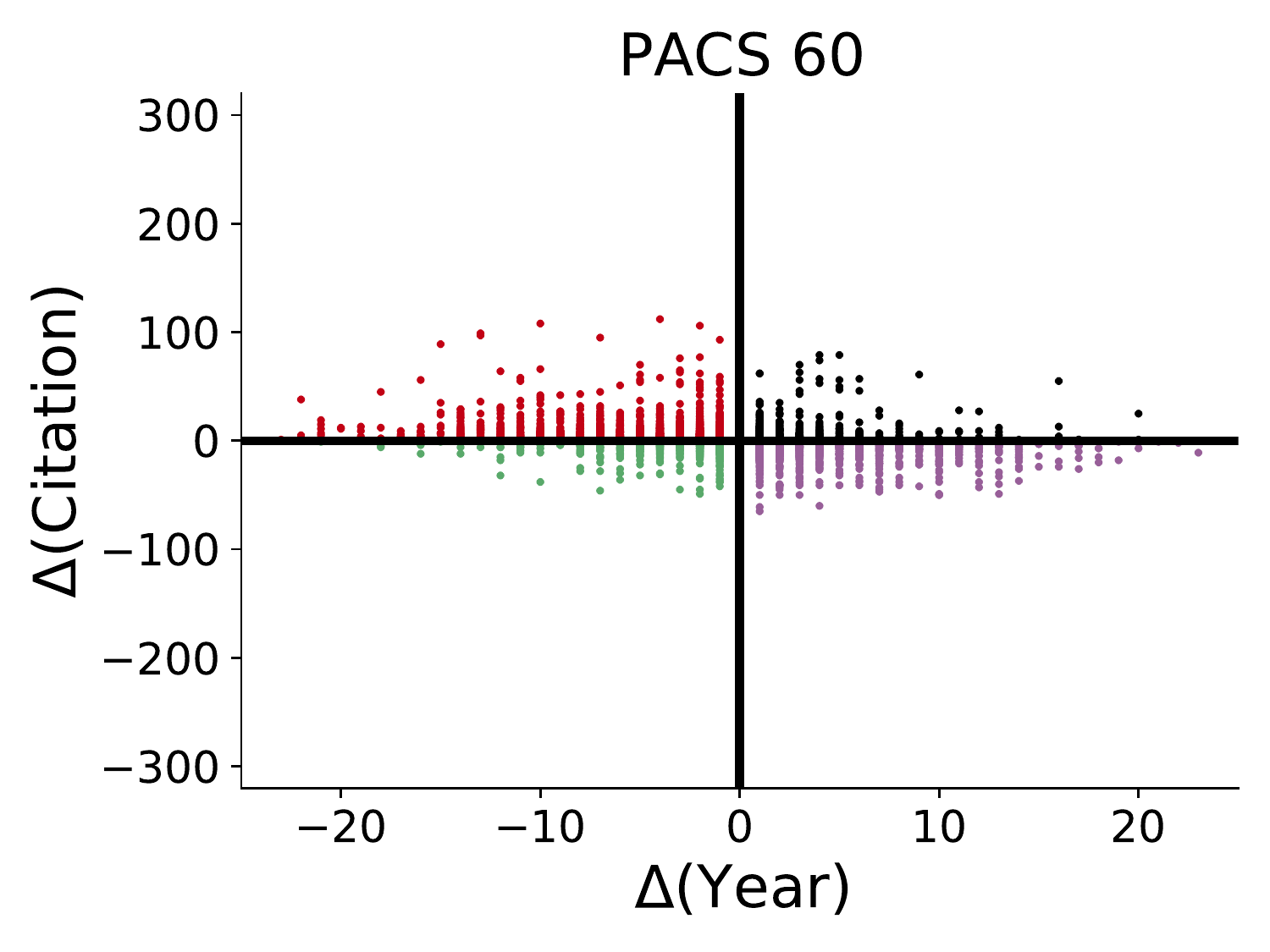}
\end{subfigure}%
\begin{subfigure}{.5\textwidth}
  \centering
  \includegraphics[width=.65\linewidth]{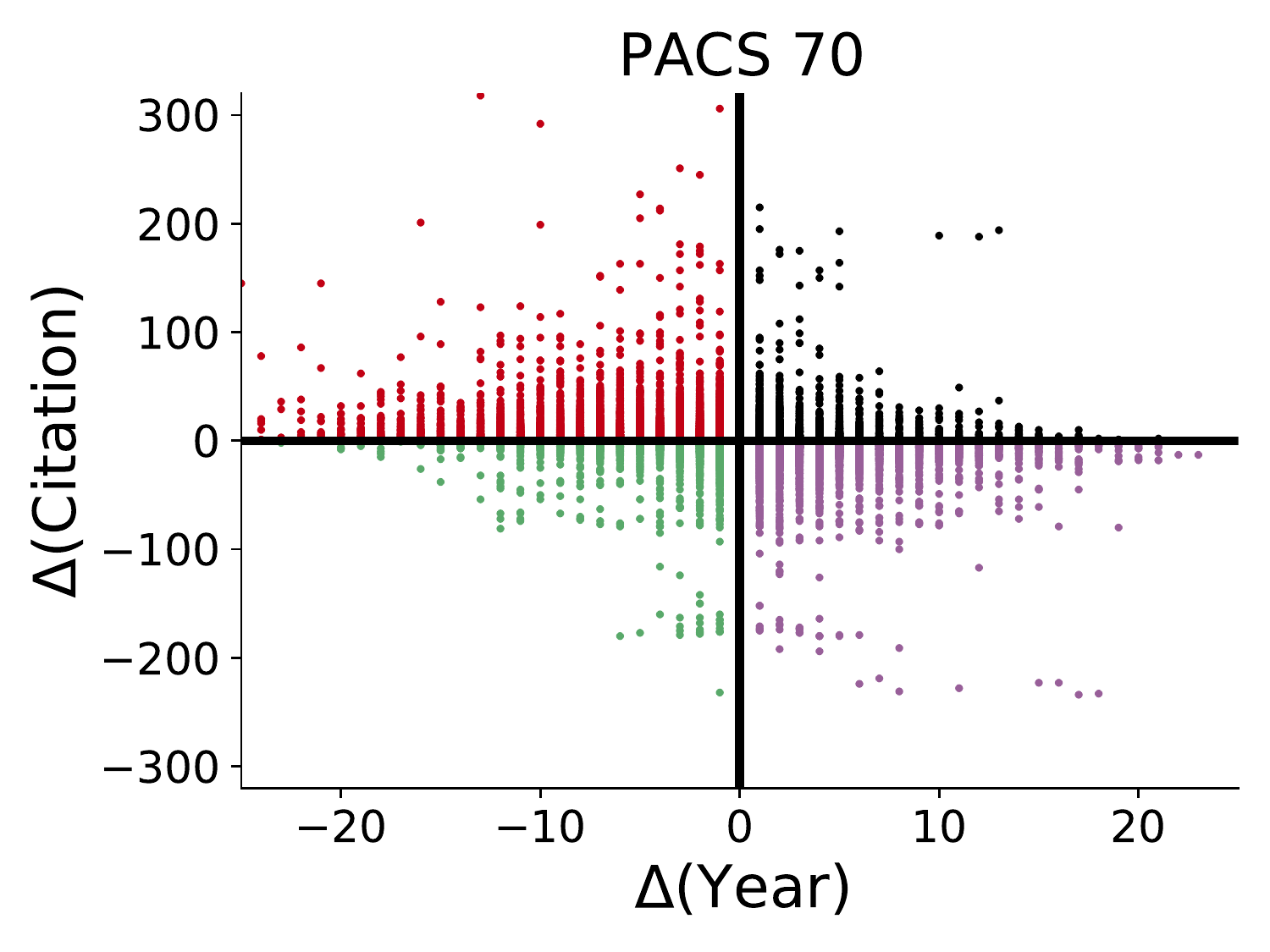}
\end{subfigure}
\begin{subfigure}{.5\textwidth}
  \centering
  \includegraphics[width=.65\linewidth]{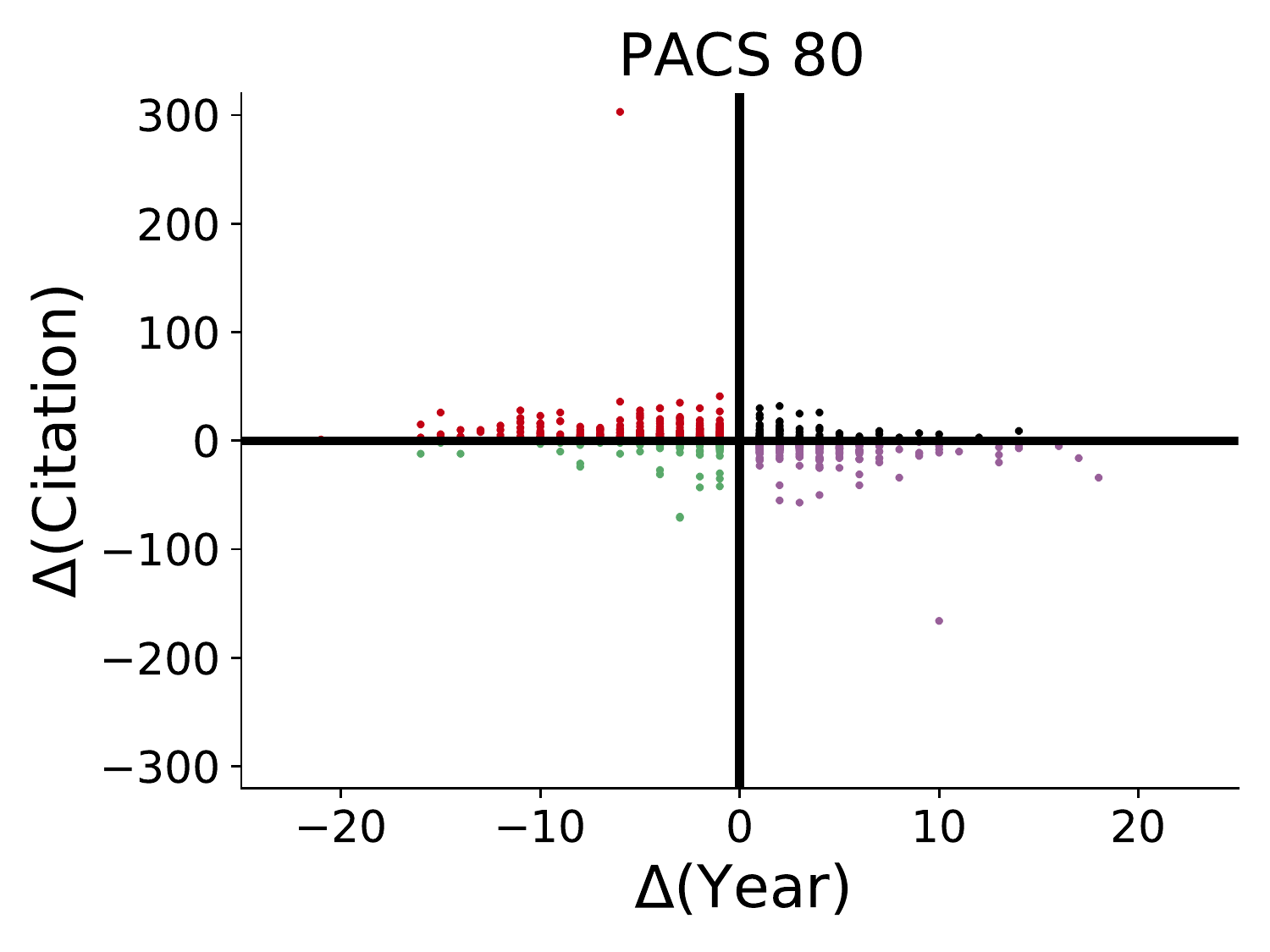}
\end{subfigure}%
\begin{subfigure}{.5\textwidth}
  \centering
  \includegraphics[width=.65\linewidth]{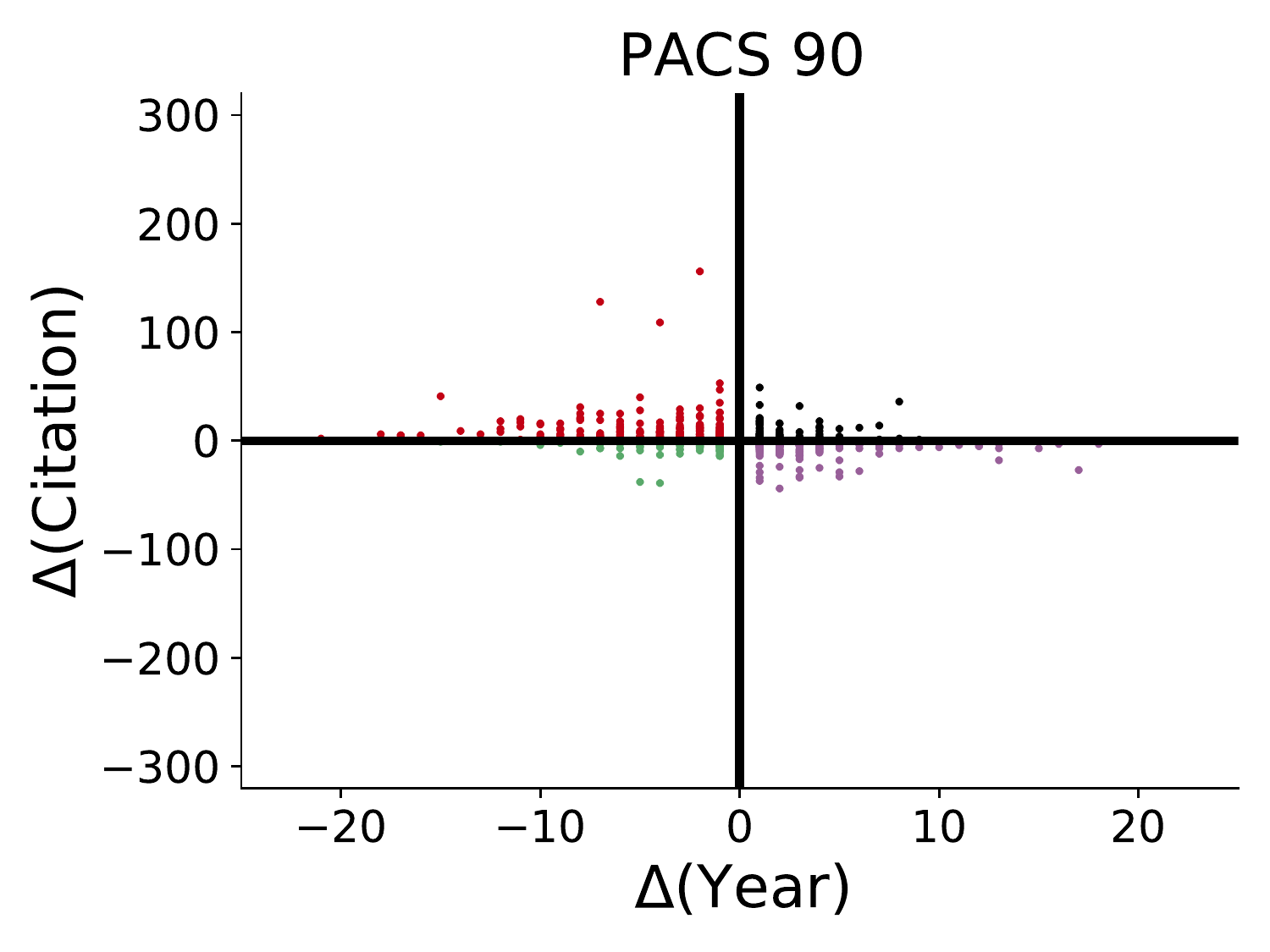}
\end{subfigure}
\caption{\textbf{Centrality and year difference for similar pairs of papers.} As in Figure 3B of the main document, these scatterplots display the centrality difference and the year difference between similar male-female pairs of papers in each subfield.}
\label{fig:cyds}
\end{figure}

 \begin{figure}[h]
        \centering
        \includegraphics[width=0.5\textwidth]{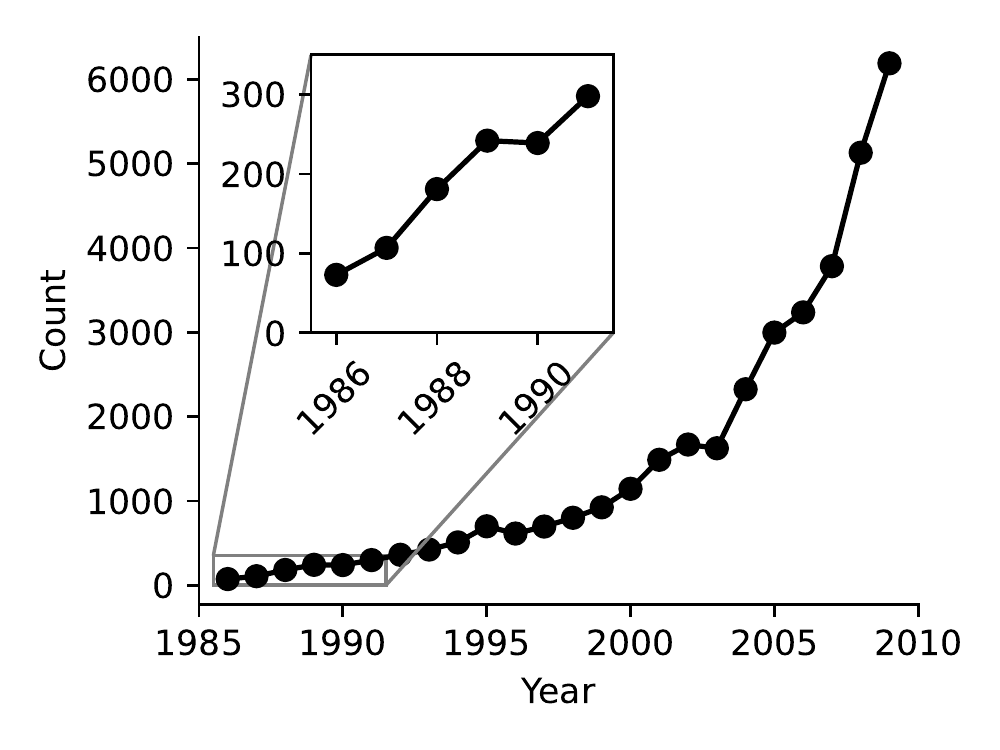}
        \caption{\textbf{Number of sampled similar pairs of papers by publication year.} This figure shows the number of sampled similar pairs for the centrality difference analysis between similar pairs per year (see Figure 3C in the main document).}
        \label{fig:simcount}
    \end{figure}

        \begin{figure}[h!]
        \centering
        \includegraphics[width=0.47\textwidth]{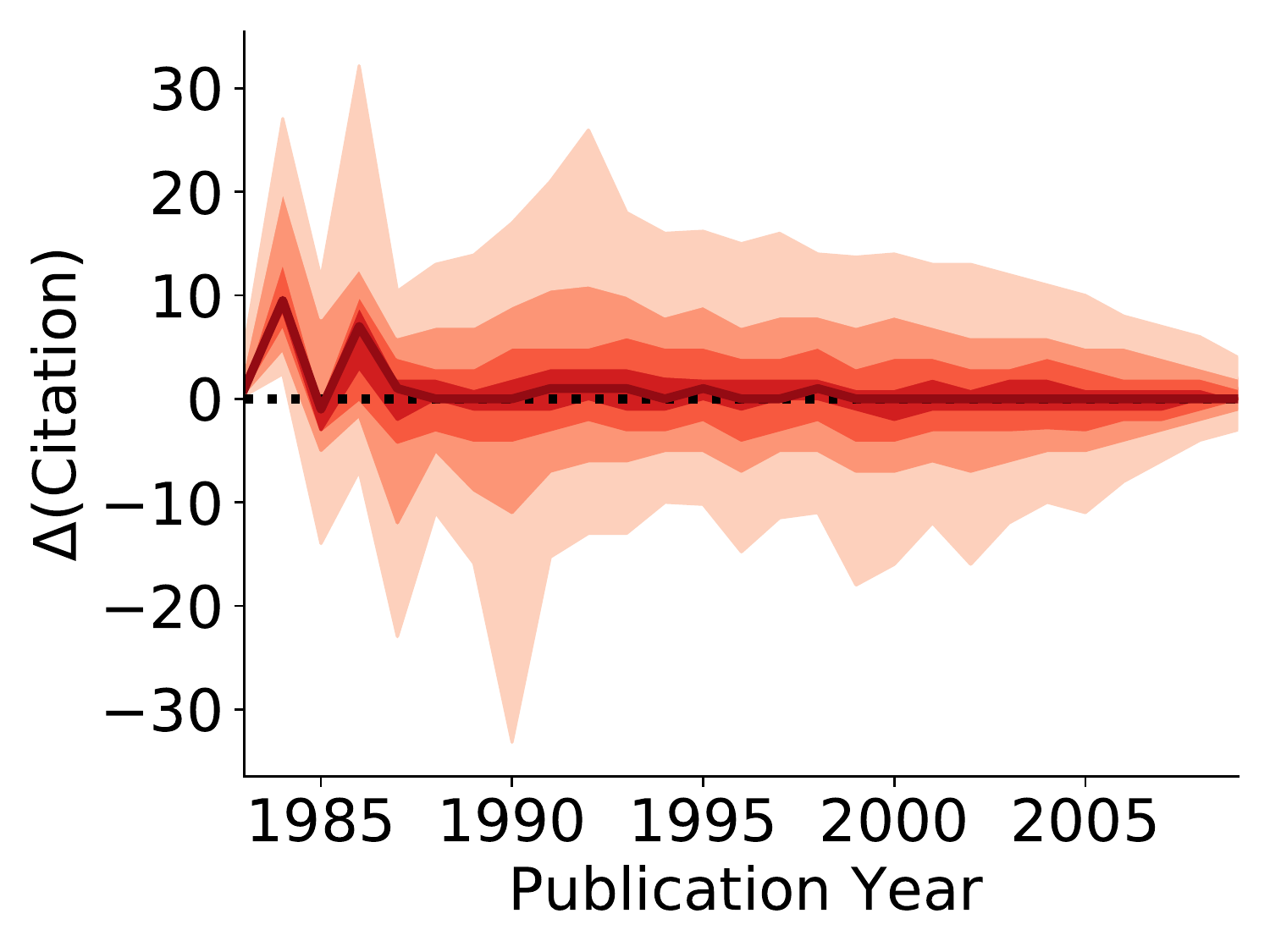}
        \includegraphics[width=0.47\textwidth]{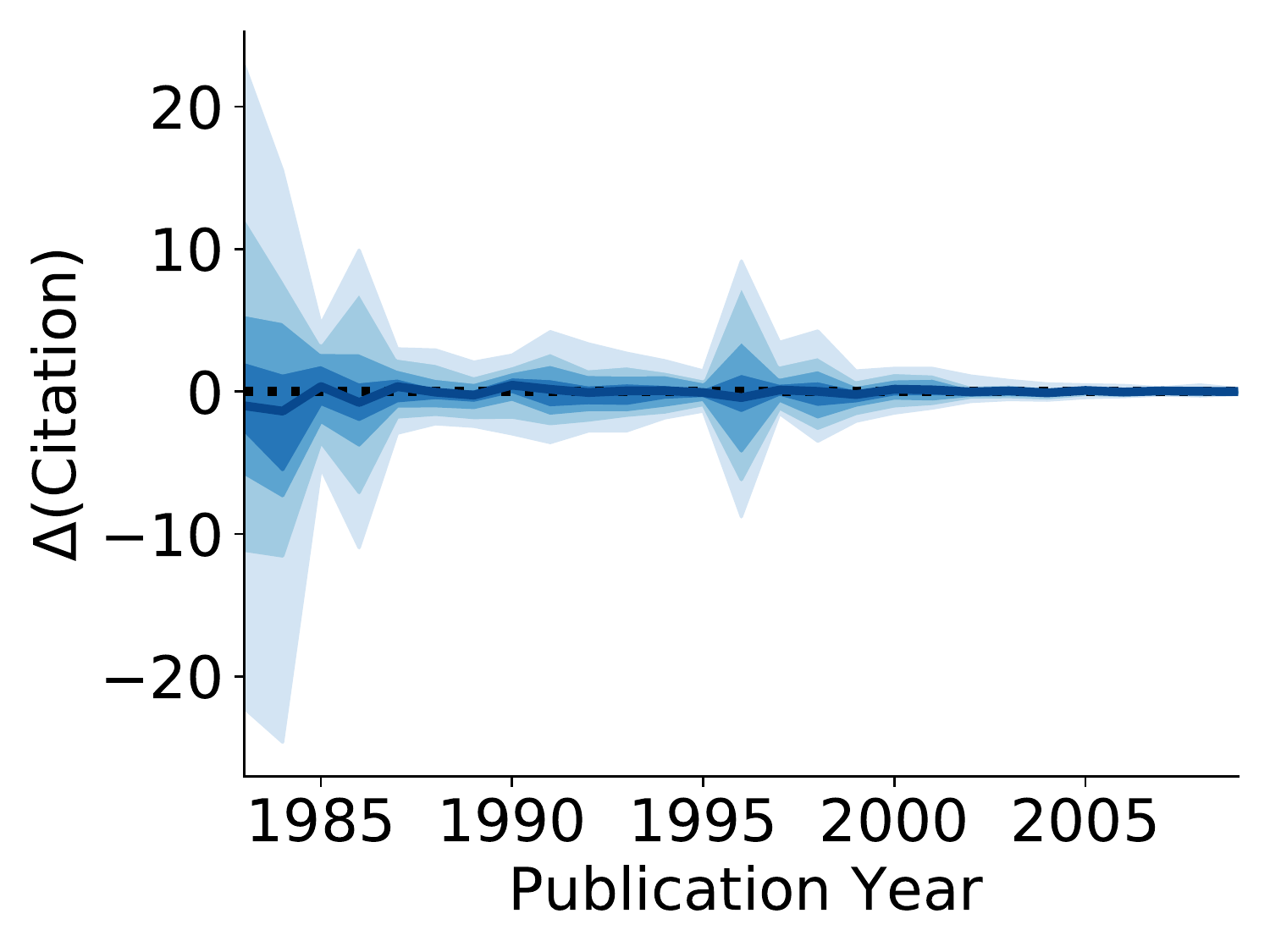}
        \caption{\textbf{Percentile plots of centrality difference by year.} Percentile plots showing the evolution of the distribution of centrality differences for similar male-female (left) and male-male (right) pairs over the years. The mean and standard errors of these distributions are shown in Figure 3C of the main document. Percentiles 10\% to 90\% are shown in different shades of red (male-female) and blue (male-male) in steps of 10\%. The two papers within each pair are published no more than 3 years from each other, and their citation difference is assigned to the year when the latter paper is published. We performed a robustness check by assigning different $p^*$ values and time intervals, and the resulting plots returned similar distributions.}
        \label{fig:pp}
    \end{figure}


\clearpage
\section{Additional supplementary tables}

\begin{table}[h]
    \centering
        \begin{tabular}{rccccc} \toprule
        \multicolumn{1}{c}{} & \textit{n} & $n_{\text{male}}$ & $p_{\text{male}}$ & $n_{\text{female}}$ & $p_{\text{female}}$ \\ \midrule
        \multicolumn{1}{l}{Total \# of citations} & 9,384,218 & 8,516,293 & 0.9075 & 867,925 & 0.0925 \\
        Self-citations & 564,630 & 524,788 & 0.9294 & 39,572 & 0.0706 \\
        Self-citation ratio & (6.02\%) & (6.16\%) &  & (4.56\%) &  \\ \hline
        \multicolumn{1}{l}{Total \# of observed authors} & 68,505 & 58,888 & 0.8596 & 9,617 & 0.1404 \\
        Self-citing authors & 36,070 & 31,987 & 0.8868 & 4,083 & 0.1132 \\
        Self-citing author ratio & (52.65\%) & (54.32\%) &  & (42.46\%) & \\        \bottomrule
        \end{tabular}
        \caption{\textbf{Proportion of self-citation and self-citing authors by gender.} This table shows the statistics on self-citations by gender. It illustrates the proportions of self-citation performed by male and female APS authors as well as the proportions of self-citing male and female authors.}
    \label{table:selfcit}
    \end{table}
    
\begin{table}[h]
    \centering
    \begin{tabular}{cccc}
    \toprule
Subfield & Total Papers & \begin{tabular}[c]{@{}c@{}}Alphabetically\\ Ordered Papers\end{tabular} & Alphabetical \% \\ \midrule
PACS 00  & 50,719       & 2,582                                                                   & 5.091           \\
PACS 10  & 24,142       & 3,218                                                                   & 13.329          \\
PACS 20  & 14,510       & 879                                                                     & 6.058           \\
PACS 30  & 23,145       & 707                                                                     & 3.055           \\
PACS 40  & 28,800       & 967                                                                     & 3.358           \\
PACS 50  & 5,610        & 146                                                                     & 2.602           \\
PACS 60  & 50,569       & 1,270                                                                   & 2.511           \\
PACS 70  & 87,066       & 1,754                                                                   & 2.015           \\
PACS 80  & 25,281       & 628                                                                     & 2.484           \\
PACS 90  & 9,222        & 902                                                                     & 9.781          \\\bottomrule
\end{tabular}
    \caption{\textbf{Proportion of alphabetically ordered papers by subfield.}}
    \label{table:ao}
\end{table}

\begin{table}[h]
    \centering
    \begin{tabular}{cccccccc}
    \toprule
    Position & $n_m + n_f$ & $n_m$ & $p_m$ & $n_f$ & $p_f$ & $z$ & $p$-value \\ \midrule
    First & 149,627 & 137,223 & 0.2986 & 12,404 & 0.2834 & \textbf{6.6346} & \textless 0.00001 \\
    Second & 87,869 & 80,073 & 0.1742 & 7,796 & 0.1781 & -2.052 & 0.9799 \\
    Middle & 115,619 & 104,827 & 0.2281 & 10,792 & 0.2466 & -8.7869 & \textgreater 0.99999 \\
    Last & 150,182 & 137,412 & 0.2990 & 12,770 & 0.2918 & \textbf{3.1535} & 0.0008 \\    \bottomrule
    \end{tabular}
    \caption{\textbf{Statistical tests for author order analysis.} In this table every pair (publication,author) is a unique data point, so each author appears repeated the number of times he or she has published in a given position. As a result, $n_f$ (resp. $n_m$) is the number of times a female (resp. male) author appears in a paper in the corresponding position. \textit{z}-scores and \textit{p}-values are accordingly calculated (see Methods) and are rounded up to the fourth decimal places with an exception of extreme values. \textit{n} and \textit{p} respectively denote sample size and proportion.}
    \label{table:aa}
    \end{table}
    
\begin{table}[h]
        \centering
        \begin{tabular}{cccccccc}
        \toprule
         & \textit{n} & $n_\text{male}$ & $p_\text{male}$ & $n_\text{female}$ & $p_\text{female}$ & \textit{z} & \textit{p}-value \\ \midrule
        \begin{tabular}[c]{@{}c@{}}Top 10\%\\ (511+ citations)\end{tabular} & 40 & 39 & 0.9750 & 1 & 0.0250 & 1.279 & 0.1004 \\
        \begin{tabular}[c]{@{}c@{}}Top 20\%\\ (346+ citations)\end{tabular} & 90 & 89 & 0.9889 & 1 & 0.0111 & 2.405 & 0.0081 \\
        \begin{tabular}[c]{@{}c@{}}Top 30\%\\ (288+ citations)\end{tabular} & 152 & 149 & 0.9803 & 3 & 0.0197 & 2.733 & 0.0031 \\ 
        \begin{tabular}[c]{@{}c@{}}Top 40\%\\ (243+ citations)\end{tabular} & 226 & 220 & 0.9735 & 6 & 0.0265 & 2.954 & 0.0016 \\        \bottomrule
        \end{tabular}
        \caption{\textbf{Statistical tests comparing degree centrality by gender in the top ranks.} 
        Comparison of the proportion of papers respectively led by male and female primary authors in the top ranks of degree centrality. For reference, the overall proportion of female led papers is 0.08. The high $z$-scores and low $p$-values corroborate the gender disparities found in Figure 3A of the main document.
        }
    \label{table:fcp}
    \end{table}
    
\begin{table}[ht]
    \centering
        \begin{tabular}{ccccccccc}
        \toprule
        PACS & Subfield & $N_{mf}$ & $p^*$ & $|M(p^*)|$ & $\frac{|M(p^*)|}{N_{mf}}$ & $d(p^*)$ & $z$ & \textit{p}-value \\ \midrule
        00 & General Physics & 184694 & 0.002 & 9931 & 5.38\% & -0.398 & -1.181 & 0.238\\ 
        10 & Elementary Particles and Fields & 49254 & 0.003 & 2833 & 5.75\% & 0.758 & \textbf{2.908} &0.0036\\ 
        20 & Nuclear Physics & 7698 & 0.003 & 385 & 5.00\% & 0.584 & 1.453 &0.146\\ 
        30 & Atomic and Molecular Physics & 29246 & 0.002 & 1474 & 5.04\% & 1.028 & \textbf{3.058} &0.0022\\ 
        40 & \begin{tabular}{@{}c@{}}Electromagnetism, Optics, Acoustics, Heat\\Transfer, Classical Mechanics, Fluid Dynamics\end{tabular}& 54621 & 0.0025 & 2525 & 4.62\% & 0.526 & 1.889 & 0.059\\ 
        50 & Gases, Plasmas, Electric Discharges & 747 & 0.006 & 48 & 6.43\% & -0.021 & -0.032 &0.974\\ 
        60 & Condensed Matter (CM): Mechanical, Thermal& 123631 & 0.0018 & 7063 & 5.71\% & 0.432 & \textbf{3.039} &0.0024\\ 
        70 & CM: Electrical, Magnetic, Optical & 529069 & 0.002 & 28952 & 5.47\% & 0.674 & \textbf{5.623} & \textless 0.00001\\ 
        80 & Interdisciplinary Physics \& Related Studies & 29173 & 0.0025 & 1602 & 5.49\% & -0.408 & -0.860 & 0.390\\ 
        90 & Geophysics, Astronomy, Astrophysics & 18760 & 0.006 & 1041 & 5.55\% & 1.603 & \textbf{4.266} & 0.00002 \\ \bottomrule
        \end{tabular}
                \caption{\textbf{Differences in received citations among similar pairs of publications.} Gender differences in received citations among pairs of publications with validated similarity measured by $z$-scores. The variables of the columns are the following (more details in Methods): $N_{mf}$ - number of all possible male-female pairs; $p^*$ - chosen critical similarity value, the lower, the more similar; $M(p^*)$ - subset of pairs with similarity of $p^*$ or better; $d(p^*)$ - average male-female citation difference; $z$ - normalized difference of male-female average citations. Values of $p^*$ are chosen to establish $\frac{|M(p^*)|}{N_{mf}}$ values between 4\% and 7\%. Significant $z$-scores are marked in bold.}
        \label{table:homa1}
    \end{table}
    
\begin{sidewaystable}[]
\centering
\resizebox{\textwidth}{!}{
\begin{tabular}{ccccccccccccccccccc}
\toprule
          & \multicolumn{2}{c}{PACS 00} & \multicolumn{2}{c}{PACS 10}        & \multicolumn{2}{c}{PACS 20} & \multicolumn{2}{c}{PACS 30} & \multicolumn{2}{c}{PACS 40}        & \multicolumn{2}{c}{PACS 60} & \multicolumn{2}{c}{PACS 70} & \multicolumn{2}{c}{PACS 80} & \multicolumn{2}{c}{PACS 90}       \\ \midrule
Year Diff & Q1 vs. Q3        & Q4 vs. Q2 & Q1 vs. Q3        & Q4 vs. Q2        & Q1 vs. Q3  & Q4 vs. Q2       & Q1 vs. Q3        & Q4 vs. Q2 & Q1 vs. Q3        & Q4 vs. Q2        & Q1 vs. Q3     & Q4 vs. Q2    & Q1 vs. Q3     & Q4 vs. Q2    & Q1 vs. Q3        & Q4 vs. Q2 & Q1 vs. Q3       & Q4 vs. Q2       \\ \midrule
1         & -1.0276          & -0.4593   & 0.8158           & \textbf{2.4389}  & -0.1137    & 1.3310          & 1.8284           & 1.8513    & 1.9239           & -1.0231          & 0.4762        & -0.6276      & -0.9009       & -0.4480      & -0.2413          & 0.0963    & \textbf{2.4604} & 1.1396          \\
2         & -0.7112          & -0.6080   & -0.4986          & 1.1457           & 0.1802     & -0.4811         & 1.3864           & 0.3983    & 0.4605           & -0.4848          & -0.6398       & 0.5378       & -0.5815       & -0.5130      & 1.2425           & -0.4183   & 0.6071          & 1.3616          \\
3         & 0.2387           & -1.2630   & 0.8695           & 0.1341           & 1.5063     & -0.7702         & -0.3992          & 0.7234    & -1.0897          & \textbf{-2.4929} & 1.3901        & -0.0954      & -0.5551       & 0.4223       & -1.0803          & -1.0713   & 0.6929          & -0.8376         \\
4         & 0.0998           & -1.4989   & -1.2089          & -0.0157          & 0.6585     & 0.3445          & 1.1498           & 0.3478    & \textbf{-2.3946} & 1.2368           & 1.2652        & 0.8284       & 0.1545        & -1.0583      & -0.6824          & -1.1105   & -0.2388         & 1.1283          \\
5         & \textbf{2.5848}  & -0.2843   & 0.8910           & 0.2543           & 0.6547     & 0.9623          & 1.1239           & 0.0808    & \textbf{2.0510}  & 1.4468           & 1.6584        & 0.1256       & 1.4723        & -0.2557      & 0.2562           & 0.7429    & -0.6698         & -1.3254         \\
6         & 0.2919           & -1.4444   & 0.6168           & -0.7551          & -          & 0.5976          & 0.7370           & 0.5413    & -1.2679          & -0.9277          & 0.4748        & -0.2807      & -1.2351       & 0.7690       & -0.7708          & 0.8244    & -               & 0.1472          \\
7         & -0.8214          & -0.2422   & 1.0674           & -0.6662          & -          & \textbf{7.5951} & 1.3540           & 0.7106    & 1.6427           & 0.1136           & -1.2621       & -1.2536      & -0.7857       & 0.9365       & -                & -1.2055   & -0.0735         & 0.9886          \\
8         & -1.9479          & -0.2470   & \textbf{-2.5156} & 0.5238           & -          & -               & -                & -0.4330   & -1.3060          & 1.8453           & -0.2936       & -0.6337      & -1.4345       & -1.1993      & \textbf{-2.2968} & -0.8736   & -               & \textbf{2.9049} \\
9         & -0.7831          & 1.1800    & 1.0545           & -1.5591          & -          & -0.9384         & \textbf{2.5156}  & -0.8575   & -                & 0.2941           & 1.0550        & 0.8075       & -1.6362       & 1.7124       & -                & 0.6258    & -               & \textbf{2.0403} \\
10        & 0.8334           & -0.2938   & 0.4264           & \textbf{-1.9832} & -          & -0.1902         & 0.2774           & -0.3513   & 1.3392           & -0.6085          & -0.4489       & -1.3741      & 0.9067        & 0.6636       & 0.8944           & -0.9810   & -               & -              \\\bottomrule
\end{tabular}
}
\caption{\textbf{Statistical tests of gender asymmetry in the first-mover advantage.} 
Comparison of the citation differences between quadrants $Q1 / Q3$ and $Q2 / Q4$ of Figure \ref{fig:cyds} for each year difference. The values shown in this table are $z$-scores computed according to equation (11) of the main document. Most of them lie in the range $(-2,2)$ (not significant), indicating that there is no gender asymmetry in the advantage gained by an author by publishing earlier. Data with less than three data points do not yield meaningful statistics and therefore are excluded from our analysis (they are marked as '-'). PACS 50 has a very small sample size and hence is not analyzed.
}
\label{table:cyd}
\end{sidewaystable}

\end{document}